\documentclass{cernyrep}
\usepackage{texnames}
\usepackage[T1]{fontenc}
\usepackage{varwidth}
\usepackage{xcolor}

\begin{document}
\title{Beam--Beam Effects}

\author{W. Herr and T. Pieloni}

\institute{CERN, Geneva, Switzerland}

\maketitle 

\begin{abstract}
One of the most severe limitations in high-intensity
particle colliders is the beam--beam interaction,
i.e. the perturbation of the beams as they cross
the opposing beams. This introduction to beam--beam effects concentrates on a description of the phenomena that are present in modern colliding beam facilities.
\end{abstract}

\section{Introduction}
The problem of the beam--beam interaction is the subject of
many studies since the introduction of the first
particle colliders.
It has been and will be one of the most important limits
to the performance and therefore attracts interest
at the design stage of a new colliding beam facility.
A particle beam is a collection of a large number of charges
and represents an electromagnetic potential for other charges.
It will therefore exert forces on itself and other beams.
The forces are most important for high-density beams, i.e. high intensities
and small beam sizes, which are the key to high luminosity.

The electromagnetic forces from particle beams are very non-linear
and result in a wide spectrum of consequences for the beam
dynamics.
Furthermore, as a result of the interaction, the charge distribution
creating the disturbing fields can change as well.
This has to be taken into account in the evaluation of beam--beam effects
and in general a self-consistent treatment is required.

Although we now have a good qualitative understanding of the
various phenomena, a complete theory does not exist and exact
predictions are still difficult.
Numerical techniques such as computer simulations have been
used with great success to improve the picture on some
aspects of the beam--beam interaction while for other problems
the available models are not fully satisfactory in their
predictive power.

\section{Beam--beam force}
In the rest frame of a beam we have only electrostatic fields
and, to find the forces on other moving charges, we have
to transform the fields into the moving frame and to
calculate the Lorentz forces (see \cite{aaa64} to \cite{aaaa5} and references therein).

The fields are obtained by integrating over the charge distributions.
The forces can be defocusing or focusing, since the test particle
can have the same or opposite charge with respect to the beam producing
the forces.

The distribution of particles producing the fields can follow
various functions, leading to different fields and forces.
It is not always possible to integrate the distribution to
arrive at an analytical expression for the forces, in which
case either an approximation or numerical methods have to be used.
This is in particular true for hadron beams, which usually do not
experience significant synchrotron radiation and damping.
For e$^{-}$e$^{+}$ colliders the distribution functions
are most likely Gaussian with truncated tails.

In the two-dimensional case of a beam with bi-Gaussian
beam density distributions in the transverse planes, i.e.
$\rho(x,y)=\rho_{x}(x)~\cdot~\rho_{y}(y)$
with r.m.s. of $\sigma_{x}$ and $\sigma_{y}$,
\begin{equation}\label{eq:003}
\ \ \displaystyle{\rho_{u}(u) = \frac{1}{\sigma_{u}\sqrt{2\pi}}
\exp{\left(-\frac{u^{2}}{2\sigma_{u}^{2}}\right)}},
\ {\mathrm{where}} \ \ u = x,y,
\end{equation}
one can give the two-dimensional potential $U(x,y,\sigma_{x},\sigma_{y})$ as a
closed expression:
\begin{equation}\label{eq:004}
U(x,y,\sigma_{x},\sigma_{y}) = \frac{n e}{4\pi\epsilon_{0}}
\int_{0}^{\infty} \frac{{\mathrm{exp}}(-\frac{x^{2}}{2\sigma^{2}_{x}+q} -\frac{y^{2}}{2\sigma^{2}_{y}+q})}{\sqrt{(2\sigma^{2}_{x} + q)(2\sigma^{2}_{y} + q})} {\mathrm d}q,
\end{equation}
where $n$ is the line density of particles in the beam, $e$ the elementary
charge and $\epsilon_{0}$ the permittivity of free space \cite{aaa52}.
From the potential one can derive the transverse fields $\vec{E}$ by
taking the gradient
$\vec{E}~=~-\nabla U(x,y,\sigma_{x},\sigma_{y})$.

\section{Elliptical beams}
For the above case of bi-Gaussian distributions
(i.e. elliptical beams with $\sigma_{x}~\neq~\sigma_{y}$)
the fields can be derived
and, for the case of $\sigma_{x}~>\sigma_{y}$, we have \cite{aaaa6}
\begin{equation}\label{eq:005}
E_{x} = \frac{n e}{2\epsilon_{0} \sqrt{2\pi (\sigma_{x}^2 - \sigma_{y}^2})}
{\mathrm{Im}} \left [ {\mathrm{erf}} \left (\frac{x + \mathrm{i}y}{\sqrt{2(\sigma_{x}^2 - \sigma_{y}^2})} \right ) - \mathrm{e}^{\left ( -\frac{x^{2}}{2\sigma_{x}^2} + \frac{y^{2}}{2\sigma_{y}^{2}} \right )} {\mathrm{erf}} \left ( \frac{x\frac{\sigma_{y}}{\sigma_{x}} + \mathrm{i}y\frac{\sigma_{x}}{\sigma_{y}}}{\sqrt{2(\sigma_{x}^2 - \sigma_{y}^2})} \right ) \right],
\end{equation}
\begin{equation}\label{eq:006}
E_{y} = \frac{n e}{2\epsilon_{0} \sqrt{2\pi (\sigma_{x}^2 - \sigma_{y}^2})}
{\mathrm{Re}} \left [ {\mathrm{erf}} \left (\frac{x + \mathrm{i}y}{\sqrt{2(\sigma_{x}^2 - \sigma_{y}^2})} \right ) - \mathrm{e}^{\left ( -\frac{x^{2}}{2\sigma_{x}^2} + \frac{y^{2}}{2\sigma_{y}^{2}} \right )} {\mathrm{erf}} \left ( \frac{x\frac{\sigma_{y}}{\sigma_{x}} + \mathrm{i}y\frac{\sigma_{x}}{\sigma_{y}}}{\sqrt{2(\sigma_{x}^2 - \sigma_{y}^2})} \right ) \right].
\end{equation}
The function erf($t$) is the complex error function
\begin{equation}\label{eq:007}
{\mathrm{erf}}(t)~=~\mathrm{e}^{-t^{2}} \left [  1 + \frac{2\mathrm{i}}{\sqrt{\pi}} \int_{0}^{t} \mathrm{e}^{z^{2}}~{\mathrm d}z\right ].
\end{equation}
The magnetic field components follow from
\begin{equation}\label{eq:008}
      B_{y}~=~-\beta_{r} E_{x}/c~~~~\mathrm{and}~~~~~B_{x}~=~\beta_{r} E_{y}/c.
\end{equation}
The Lorentz force acting on a particle with charge $q$ is finally
\begin{equation}\label{eq:009}
      \vec{F}~=~q (\vec{E} + \vec{v} \times \vec{B}).
\end{equation}

\section{Round beams}
With the simplifying assumption of round
beams ($\sigma_{x} = \sigma_{y} = \sigma$),
one can re-write (\ref{eq:009}) in cylindrical coordinates:
\begin{equation}\label{eq:009a}
      \vec{F}~=~q (E_{r}  +  \beta c B_{\Phi}) \times \vec{r}.
\end{equation}
From (\ref{eq:004}) and with $r^{2}~=~x^{2}~+~y^{2}$ one can
immediately write the fields from (\ref{eq:009a}) as
\begin{equation}\label{eq:009b}
 E_{r} = -\frac{n e}{4 \pi \epsilon_{0}} \cdot \frac{\delta}{\delta r} \int_{0}^{\infty} \frac{{\mathrm{exp}}(-\frac{r^{2}}{(2 \sigma^{2} + q)})}{(2 \sigma^{2} + q)} {\mathrm d}q
\end{equation}
and
\begin{equation}\label{eq:009c}
 B_{\Phi} = -\frac{n e \beta c \mu_{0}}{4 \pi} \cdot \frac{\delta}{\delta r} \int_{0}^{\infty} \frac{{\mathrm{exp}}(-\frac{r^{2}}{(2 \sigma^{2} + q)})}{(2 \sigma^{2} + q)} {\mathrm d}q.
\end{equation}
We find from (\ref{eq:009b}) and (\ref{eq:009c}) that
the force (\ref{eq:009a}) has only a radial component.
The expressions (\ref{eq:009b}) and (\ref{eq:009c}) can easily be evaluated
when the derivative is done first and ${1}/({2 \sigma^{2} + q})$
is used as integration variable.
We can now express the radial force in a closed form (using $\epsilon_{0} \mu_{0} = c^{-2}$):
\begin{equation}\label{eq:010}
F_{{{r}}}(r) = -\frac{ne^{2}(1+\beta^{2})}{2\pi\epsilon_{0}{}}
\cdot\frac{1}{r}\cdot\left[ 1 - {\mathrm{exp}}\left(-\frac{{{r^{2}}}}{2\sigma^{2}}\right)\right]
\end{equation}
and, for the Cartesian components in the two transverse planes, we get
\begin{equation}\label{eq:011}
F_{{{x}}}(r) = -\frac{ne^{2}(1+\beta^{2})}{2\pi\epsilon_{0}{}}
\cdot\frac{x}{r^{2}}\cdot\left[ 1 - {\mathrm{exp}}\left(-\frac{{{r^{2}}}}{2\sigma^{2}}\right)\right]
\end{equation}
and
\begin{equation}\label{eq:012}
F_{{{y}}}(r) = -\frac{ne^{2}(1+\beta^{2})}{2\pi\epsilon_{0}{}}
\cdot\frac{y}{r^{2}}\cdot\left[ 1 - {\mathrm{exp}}\left(-\frac{{{r^{2}}}}{2\sigma^{2}}\right)\right].
\end{equation}

\begin{figure}[htb]
\centering{
\includegraphics*[width= 14.2cm, height=7.2cm]{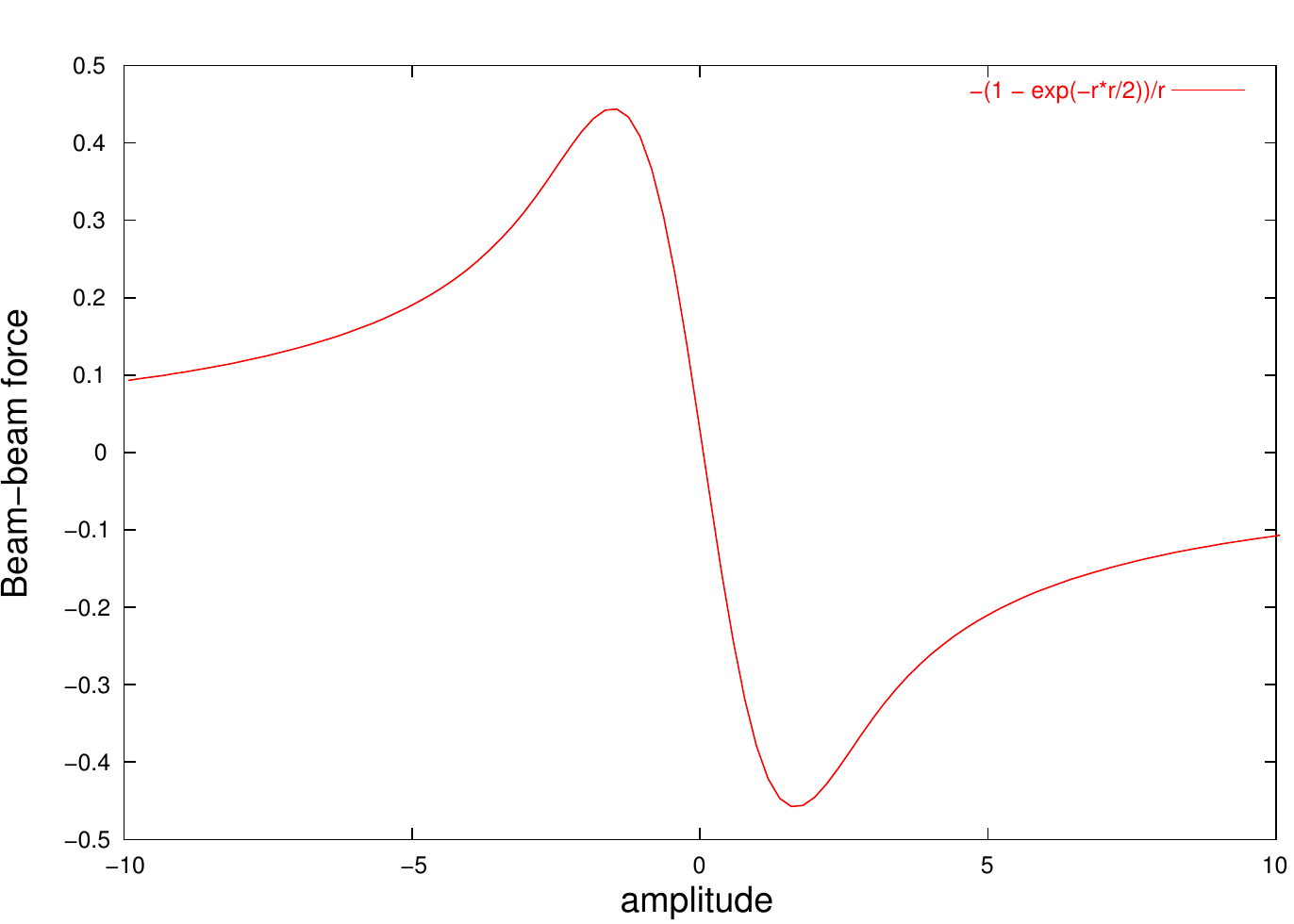}}
\caption{Beam--beam force for round beams. Force in arbitrary units, amplitude in units of r.m.s. beam size}
\label{fig:02}
\end{figure}

The forces (\ref{eq:011}) and (\ref{eq:012}) are computed when the charges of the test particle and the opposing
beam have opposite signs.
For equally charged beams the forces change sign.
The shape of the force as a function of the amplitude is given in Fig.~\ref{fig:02}.

For small amplitudes the force is approximately linear and a particle
crossing a beam at small amplitudes will experience a linear field.
This results in a change of the tune as in a quadrupole.
At larger amplitudes (i.e. above $\approx1\sigma$) the force
deviates strongly from this linear behaviour.
Particles at larger amplitudes will also experience a tune change;
however, this tune change will depend on the amplitude.
From the analytical form (\ref{eq:012}) one can see that the
beam--beam force includes higher multipoles.

\section{Incoherent effects---single-particle effects}
The force we have derived is the force of a beam on a single
test particle.
It can be used to study single-particle or incoherent effects.
For that, we treat a particle crossing a beam as if it was
moving through a static electromagnetic lens.
We have to expect all effects that are known from resonance
and non-linear theory, such as:
\begin{itemize}
\item[$\bullet$] unstable and/or irregular motion;
\item[$\bullet$] beam blow up or bad lifetime.
\end{itemize}

\section{Beam--beam parameter}
We can derive the linear tune shift of a small-amplitude
particle crossing a round beam of a finite length.
We use the force to calculate the
kick it receives from the opposing beam,
i.e. the change of the slope of the particle trajectory.
Starting from the two-dimensional force and multiplying with the longitudinal
distribution which depends on both position $s$ and time $t$ and which we
assume has a Gaussian shape with a width of $\sigma_{s}$, we obtain
\begin{eqnarray*}
F_{r}(r,s,t) = -\frac{Ne^{2}(1+\beta^{2})}{\sqrt{(2\pi)^{3}}\epsilon_{0}\sigma_{s}}
\cdot\frac{1}{r}\cdot\left[ 1 - {\mathrm{exp}}\left(-\frac{{r^{2}}}{2\sigma^{2}}\right)\right] \cdot \left[{\mathrm{exp}}\left(-\frac{{(s + vt)^{2}}}{2\sigma_{s}^{2}}\right)\right].
\end{eqnarray*}
Here $N$ is the total number of particles.
We make use of Newton's law and integrate over the collision to get the radial deflection:
\begin{eqnarray*}
\Delta r' = \frac{1}{m c \beta \gamma}\int^{\infty}_{-\infty} F_{r}(r,s,t) {\mathrm d}t.
\end{eqnarray*}
The radial kick $\Delta r'$ a particle with a radial distance $r$ from the
opposing beam centre receives is then
\begin{equation}\label{eq:013}
  \Delta r'~~=~~-\frac{2N r_{0}}{\gamma } \cdot\frac{1}{r}\cdot\left [  1 - {\mathrm{exp}} \left (  -\frac{r^{2}}{2\sigma^{2}}\right ) \right ],
\end{equation}
where we have re-written the constants and used the classical particle radius:
\begin{equation}\label{eq:014}
  r_{0} = e^{2}/4\pi \epsilon_{0} m c^{2},
\end{equation}
where $m$ is the mass of the particle.
For small amplitudes $r$ one can derive the asymptotic limit:
\begin{equation}\label{eq:015}
  \Delta r'|_{r\rightarrow 0}~~=~~-\frac{N r_{0}r}{\gamma \sigma^{2}}~~=~~-r~\cdot~f.
\end{equation}
This limit is the slope of the force at $r = 0$ and the force becomes
linear with a focal length as the proportionality factor.

It is well known how the focal length relates to a tune change
and one can derive a quantity $\xi$, which is known as the {\em linear
beam--beam parameter}:
\begin{equation}\label{eq:016}
\xi~=~\frac{N r_{0} \beta^{*}}{4\pi\gamma\sigma^{2}}.
\end{equation}
Here $r_{0}$ is the classical particle radius (e.g. $r_{\rm e}, r_{\rm p}$) and $\beta^{*}$
is the optical amplitude function ($\beta$-function) at the interaction point.

For small values of $\xi$ and a tune far enough away from linear
resonances this parameter is equal to the linear tune shift $\Delta Q$.

The beam--beam parameter can be generalized for the case of non-round beams
and becomes
\begin{equation}\label{eq:017}
{{\xi_{x,y}}}~=~\frac{N r_{0} \beta^{*}_{x,y}}{2\pi\gamma\sigma_{x,y}(\sigma_{x} + \sigma_{y})}.
\end{equation}
The beam--beam parameter is often used to quantify the strength of the
beam--beam interaction; however, it does not reflect the non-linear
nature.

\section{Non-linear effects}
Since the beam--beam forces are strongly non-linear, the study of beam--beam
effects encompasses the entire field of non-linear dynamics~\cite{herrnl} as well as collective effects.

First we briefly discuss the immediate effect of the non-linearity
of the beam--beam force on a single particle.
It manifests as an amplitude-dependent tune shift and for a beam with many
particles as a tune spread.
The instantaneous tune shift of a particle when it
crosses the other beam is related
to the derivative of the force with respect
to the amplitude $\delta F/\delta x$.
For a particle performing an oscillation with a given amplitude
the tune shift is calculated by averaging the slopes of the force
over the range (i.e. the phases) of the particle's oscillation amplitudes.

\begin{figure}[htb]\centering
\includegraphics*[height=7.2cm,width=14.2cm]{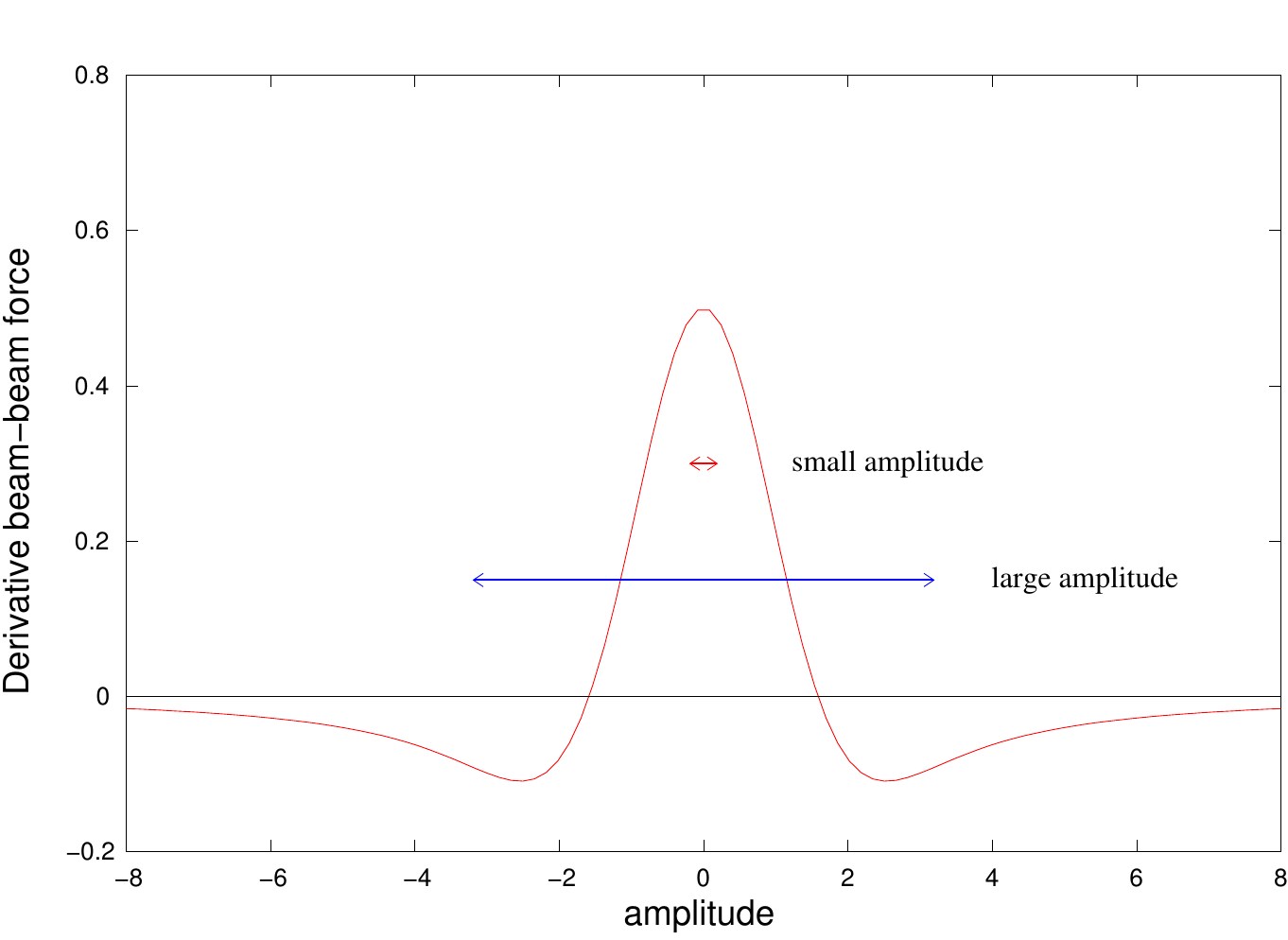}
\caption{Derivative of beam--beam force for round beams. Oscillation range of particles with large and small amplitudes.}\label{fig:02a}
\end{figure}

The derivative of the beam--beam force from Fig.~\ref{fig:02} is plotted
for the one-dimensional case in Fig.~\ref{fig:02a}.

\begin{figure}[htb]
\centering{
\includegraphics*[width= 7.5cm,height=7.2cm]{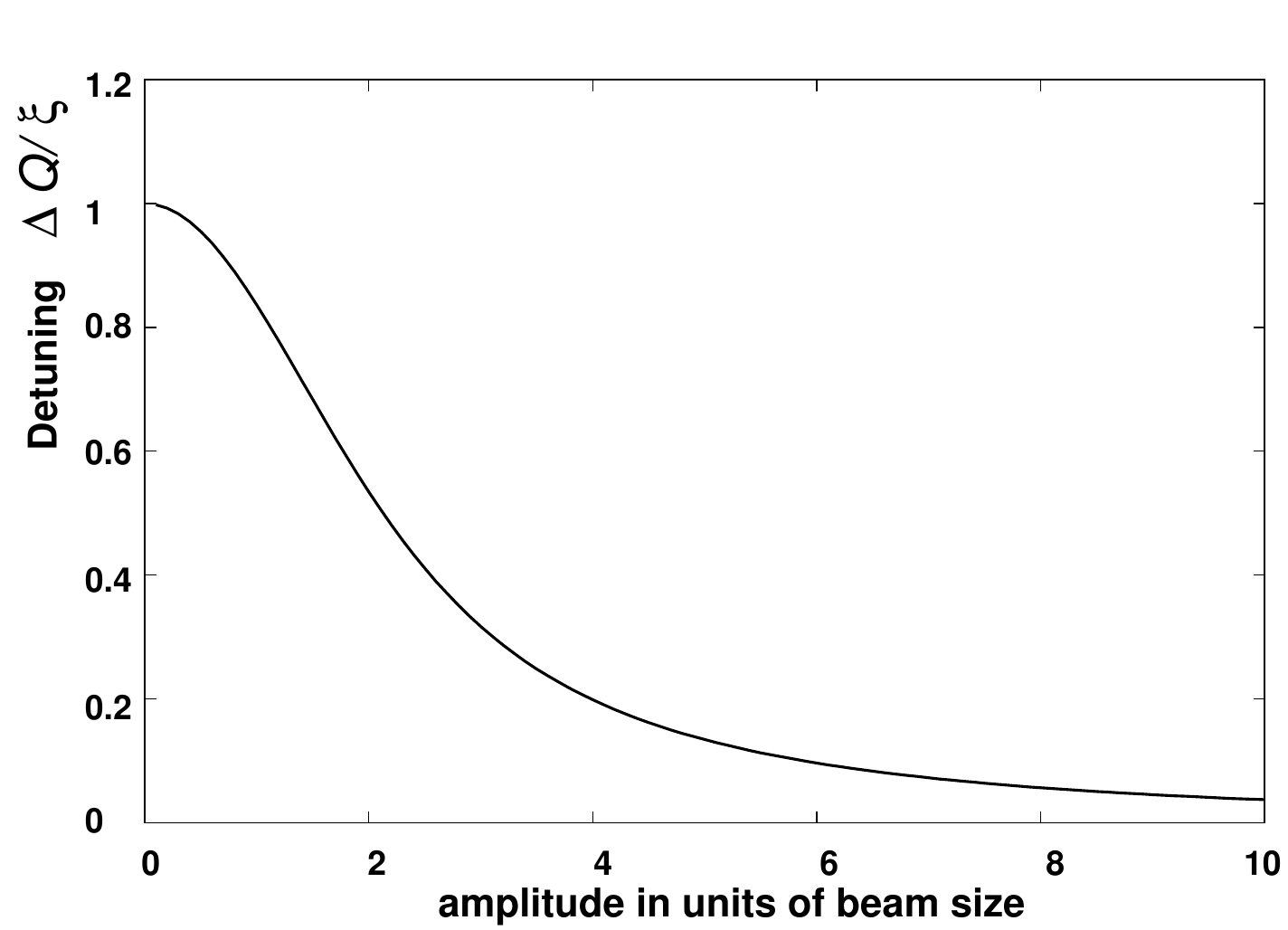}
\includegraphics*[width= 7.5cm,height=7.2cm]{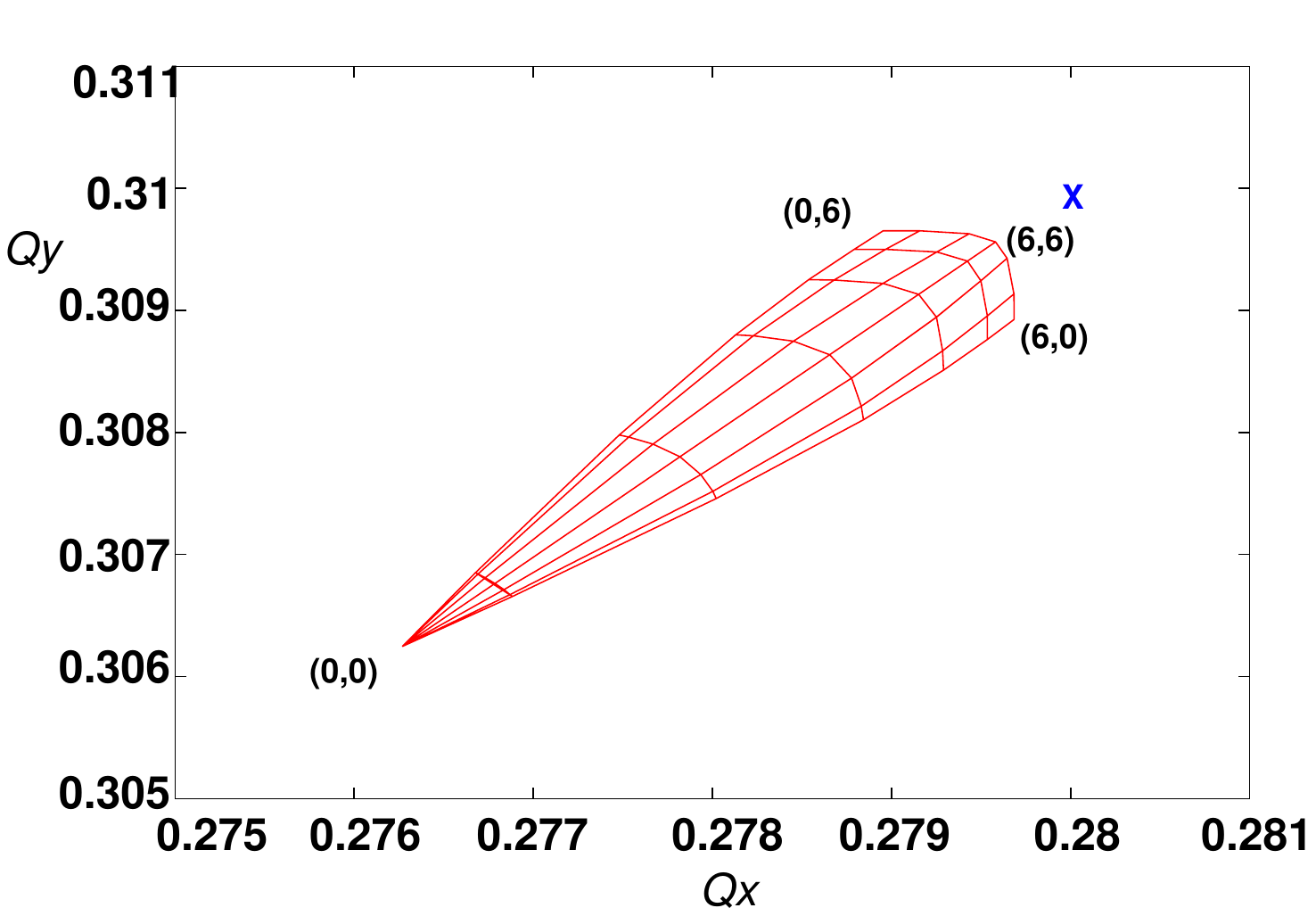}}
\caption{Tune shift (non-linear detuning) as a function of the amplitude (left) and two-dimensional tune footprint (right).}
\label{fig:03}
\end{figure}

The derivation of the detuning is given in Appendix~B, using a classical method.
An elegant calculation can be done using the Hamiltonian
formalism \cite{aaaa3} developed for non-linear dynamics and
as discussed in \cite{herrnl} using the Lie formalism.
It is demonstrated in Appendix~C for one and two interaction points and includes
the computation of the invariant Hamiltonian.
We get the formula for the non-linear detuning with the amplitude $J$:
\begin{equation}\label{eq:0b22}
\Delta Q(J)~=~\xi~\cdot~\frac{2}{J}~\cdot~( 1 - I_{0}(J/2) \cdot \mathrm{e}^{-J/2}),
\end{equation}
where $I_{0}(x)$ is the modified Bessel function
and $J~=~\epsilon \beta/2\sigma^{2}$ in the usual units.
Here $\epsilon$ is the particle `emittance' and not the beam emittance.

In the two-dimensional case, the tune shifts ($\Delta Q_{x}$, $\Delta Q_{y}$)
of a particle with amplitudes
$x$ and $y$ depend on both horizontal and vertical amplitudes.
The detuning must be computed and presented in a two-dimensional
form, i.e. the amplitude ($x$, $y$) is mapped into the tune space
($Q_{x}$, $Q_{y}$) or alternatively to the two-dimensional
tune change ($\Delta Q_{x}$, $\Delta Q_{y}$).
Such a presentation is usually called a `tune footprint' and an
example is shown in Fig.~\ref{fig:03} (right);
it maps the amplitudes into the tune space and each `knot'
of the mesh corresponds to a pair of amplitudes.
Amplitudes between 0 and 6$\sigma$ in both planes are used.
The cross indicates the original, unperturbed tunes without
the beam--beam interaction.

The maximum tune spread for a single head-on collision is
equal to the tune shift of a particle with small amplitudes and
for small tune shifts is equal to the beam--beam parameter $\xi$.
In the simple case of a single head-on collision the parameter $\xi$
is therefore a measure for the tune spread in the beam.

\section{Beam stability}
When the beam--beam interaction becomes too strong,
the beam can become unstable or the beam dynamics
is strongly distorted.
One can distinguish different types of distortions,
and a few examples are:
\begin{itemize}
\item[$\bullet$]non-linear motion can become stochastic and can result in a reduction of the dynamic aperture and particle loss and bad lifetime;
\item[$\bullet$]distortion of beam optics: dynamic beta (LEP) \cite{aaaa1};
\item[$\bullet$]vertical blow up above the so-called beam--beam limit.
\end{itemize}
Since the beam--beam force is very non-linear, the motion can
become `chaotic'.
This often leads to a reduction of the available dynamic aperture.
The dynamic aperture is the maximum amplitude where the beam
remains stable.
Particles outside the dynamic aperture can eventually get lost.
The dynamic aperture is usually evaluated by tracking particles
with a computer program through the machine where they
experience the fields from the machine elements and other
effects such as wake fields or the beam--beam interaction.

Since the beam--beam interaction is basically a very non-linear
lens in the machine, it distorts the optical properties
and it may create a noticeable beating of the
$\beta$-function around the whole machine and at the location
of the beam--beam interaction itself.
This can be approximated by inserting a quadrupole which produces
the same tune shift at the position of the beam--beam interaction.
The r.m.s. beam size at the collision point is now proportional
to $\sqrt{\beta_{\rm p}^{*}}$, where $\beta_{\rm p}^{*}$ is the perturbed
$\beta$-function, which can be significantly different from
the unperturbed $\beta$-function $\beta^{*}$.
This in turn changes the strength of the
beam--beam interaction and the parameters have to be found
in a self-consistent form.
This is called the dynamic beta effect. This is a first deviation
from our assumption that the beams are static non-linear lenses.
A strong dynamic beta effect was found in LEP \cite{aaa62} due
to its very large tune shift parameters.

Another effect that can be observed in particular in
e$^{+}$e$^{-}$ colliders is the blow up of the emittance, which
naturally limits the reachable beam--beam tune shifts.

\section{Beam--beam limit}
\begin{figure}[htb]\centering
\includegraphics*[width=10.5cm,height=7.0cm]{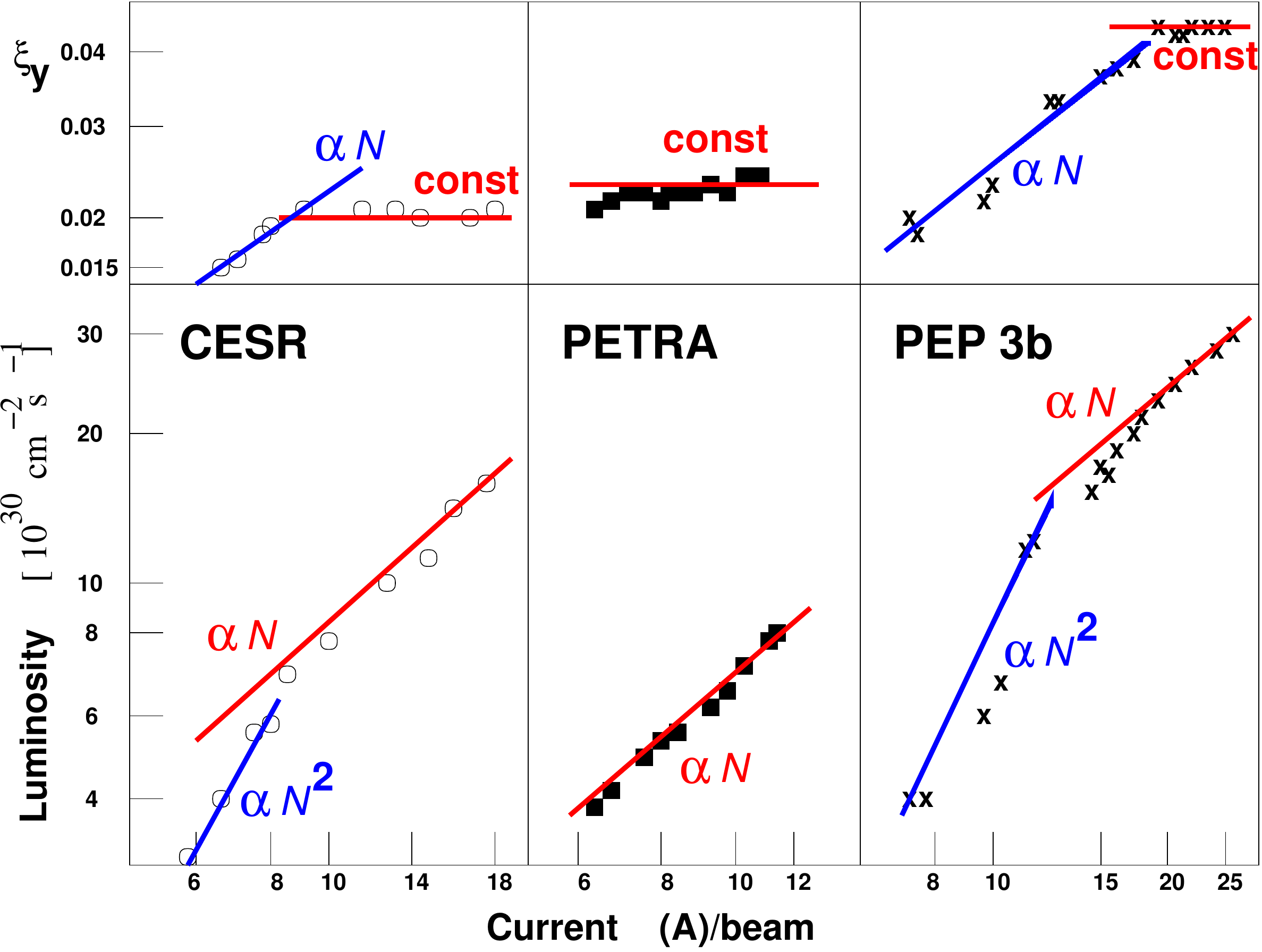}
\caption{Measurements of luminosity and beam--beam limit in e$^{+}$e$^{-}$ colliders. Logarithmic scale of the axes to demonstrate change of exponent.}\label{fig:06a}
\end{figure}

\begin{figure}[htb]\centering
\includegraphics*[width=10.5cm,height=7.0cm]{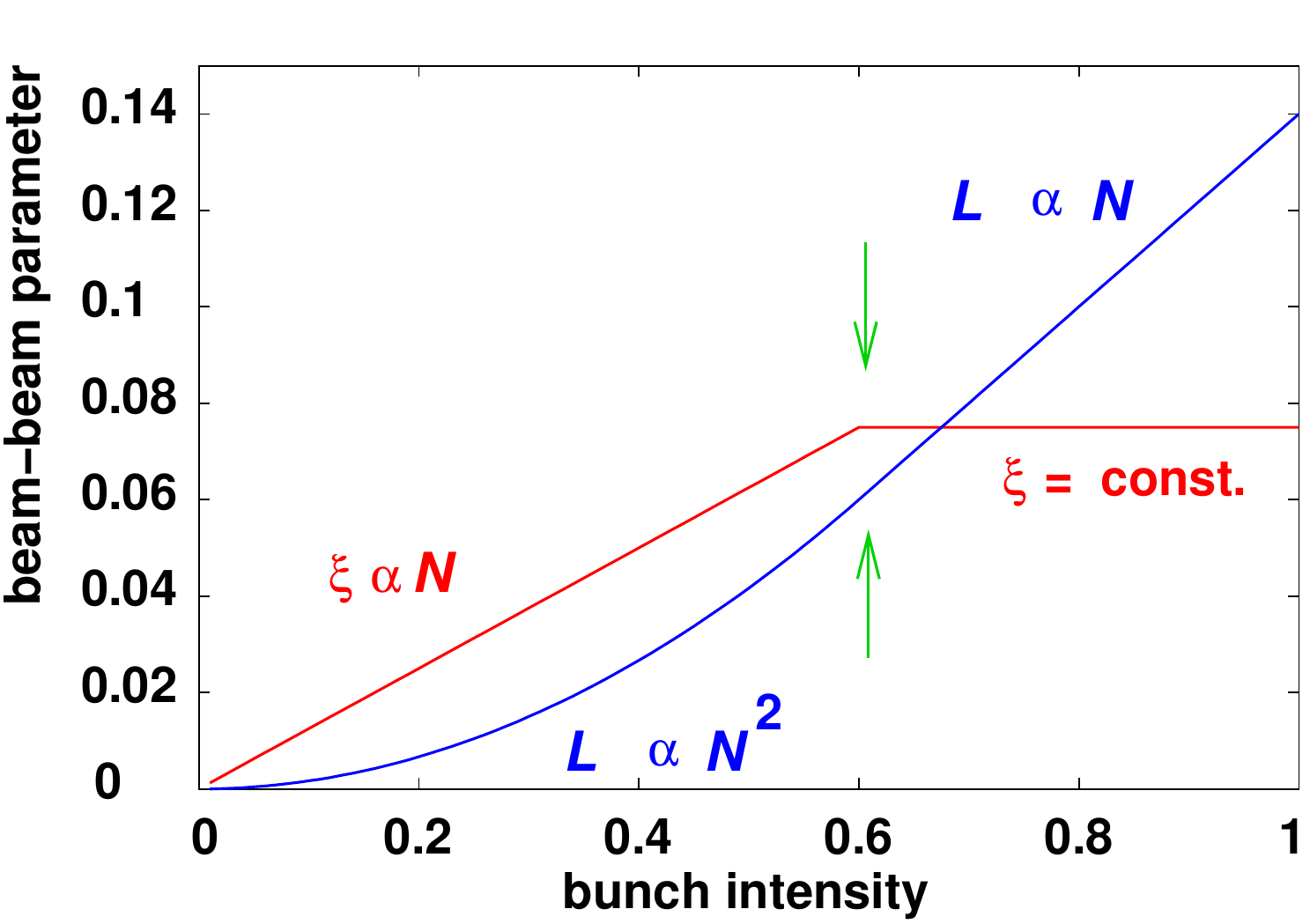}
\caption{Schematic illustration of beam--beam limit in e$^{+}$e$^{-}$ colliders}\label{fig:06b}
\end{figure}

In e$^{+}$e$^{-}$ colliders the beam sizes are usually
an equilibrium between the damping due to the synchrotron
radiation and heating mechanisms such as quantum excitation,
intra-beam scattering and, very importantly, the beam--beam effect.
This leads to a behaviour that is not observed in a hadron
collider.
When the luminosity is plotted as a function of the beam intensity,
it should increase approximately as the current squared \cite{aaa63},
in agreement with
\begin{equation}\label{eq:018}
{\cal{L}}~=~\frac{{{N^{2}}}\cdot k f}{4\pi {{\sigma_{x} \sigma_{y}}}}.
\end{equation}
Here $k$ is the number of bunches per beam and $f$ the revolution
frequency \cite{aaa63}.
At the same time the beam--beam parameter $\xi$ should increase linearly
with the beam intensity according to (\ref{eq:017}):
\begin{equation}\label{eq:019}
\xi_{y}~=~\frac{{{N}}\cdot r_{\rm e} \beta_{y}}{2\pi\gamma{{\sigma_{y}(\sigma_{x} + \sigma_{y})}}}.
\end{equation}
In all e$^{+}$e$^{-}$ colliders the observation can be made that above
a certain current, the luminosity increases approximately
proportionally to the current, or at least much less than
with the second power \cite{seeman}.
Another observation is  that at the same value of the intensity
the beam--beam parameter $\xi$ saturates.
This is shown for three e$^{+}$e$^{-}$ colliders in Fig.~\ref{fig:06a}
and schematically illustrated in Fig.~\ref{fig:06b}.
This limiting value of $\xi$ is commonly known as the {\em beam--beam limit}.


When we re-write the luminosity as
\begin{equation}\label{eq:020}
{\cal{L}}~=~\frac{{{N^{2}}}\cdot k f}{4\pi {{\sigma_{x} \sigma_{y}}}}~=~\frac{N\cdot k f}{4\pi {\sigma_{x}}} \cdot {{\frac{N}{\sigma_{y}}}},
\end{equation}
we get an idea of what is happening.
In e$^{+}$e$^{-}$ colliders the horizontal beam size $\sigma_{x}$ is usually much
larger than the vertical beam size $\sigma_{y}$ and changes very little.
In order for the luminosity to increase proportionally  to the
number of particles $N$, the factor $N/\sigma_{y}$ must be constant. 

This implies that with increasing current the vertical beam size
increases in proportion above the beam--beam limit.
This has been observed in all e$^{+}$e$^{-}$ colliders and, since the vertical
beam size is usually small, this emittance growth can be very substantial
before the lifetime of the beam is affected or beam losses are observed.

The dynamics of machines with high synchrotron radiation is dominated
by the damping properties and the beam--beam limit is not a universal
constant nor can it be predicted.
Simulation of beams with many particles can provide an idea of the
order of magnitude \cite{steve1, steve2}.

\section{Crossing angle}
To reach the highest luminosity, it is desirable to operate a collider
with as many bunches as possible, since the luminosity is proportional
to their number \cite{aaa63}.

In a single-ring collider such as the SPS, Tevatron or LEP, the operation
with $k$ bunches leads to 2$\cdot k$ collision points.
When $k$ is a large number, most of them are unwanted and must be avoided
to reduce the perturbation due to the beam--beam effects.

\begin{figure}[htb]\centering
\includegraphics*[height=6.1cm,width= 7.5cm]{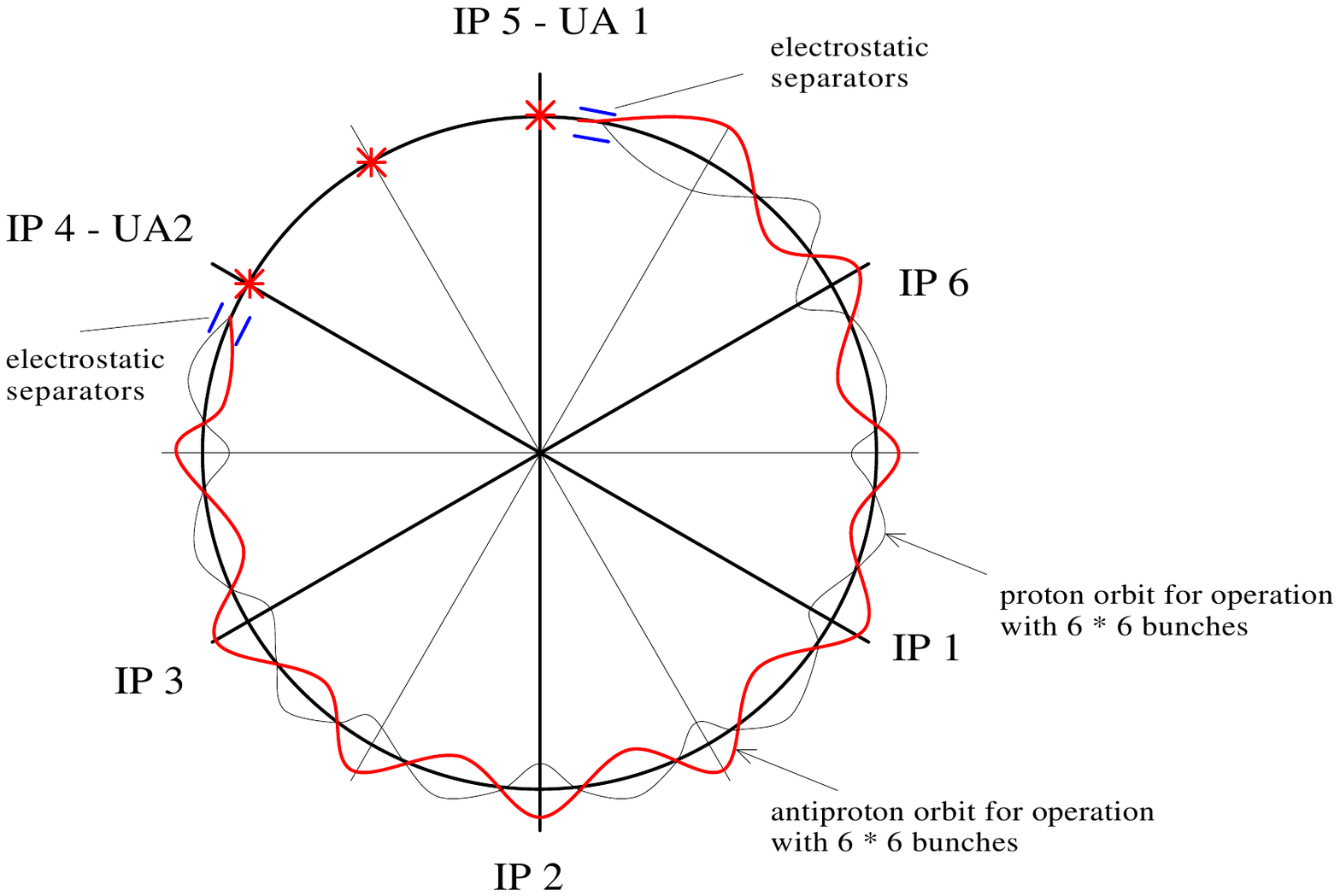}
\includegraphics*[height=6.8cm,width= 7.5cm]{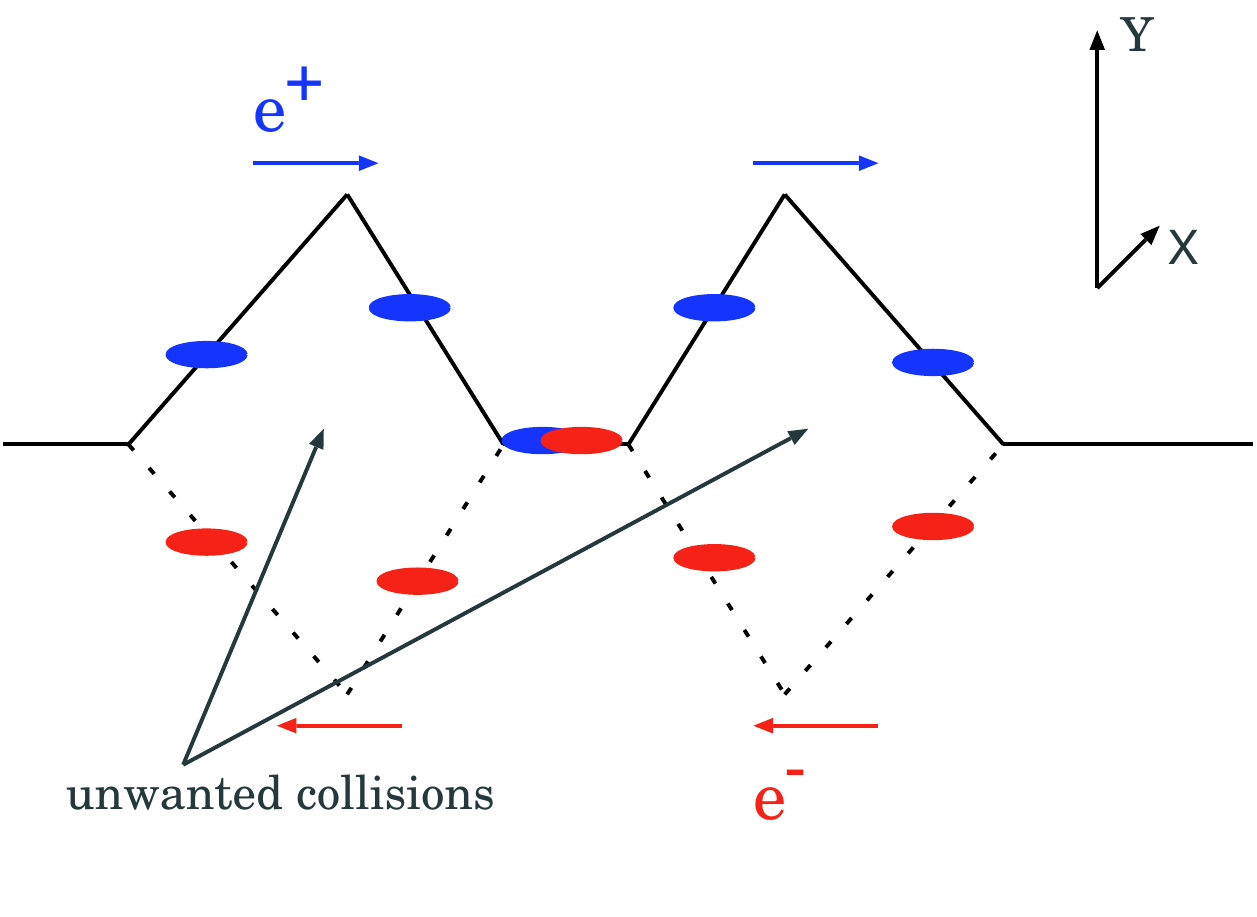}
\caption{Beam separation with a Pretzel scheme (SPS, Tevatron, LEP, left) and with short bunch trains (LEP, right)}\label{fig:012}
\end{figure}

Various schemes have been used to avoid these unwanted `parasitic' interactions.
Two prominent examples are shown in Fig.~\ref{fig:012}.
In the SPS, Tevatron and LEP so-called Pretzel schemes were used.
When the bunches are equidistant, this is the most promising method.
When two beams of opposite charge travel in the same beam pipe, they can
be moved onto separate orbits using electrostatic separators.
In a well-defined configuration the two beams cross when the beams are
separated (Fig.~\ref{fig:012}, left).
To avoid a separation around the whole machine, the bunches can be arranged
in so-called trains of bunches following each other closely.
In that case a separation with electrostatic separators is only needed
around the interaction regions.
Such a scheme was used in LEP in the second phase \cite{btlep} and it is
schematically illustrated in Fig.~\ref{fig:012} (right).

Contrary to the majority of the colliders, the LHC collides particles of
the same type, which therefore must travel in separate beam pipes.

\begin{figure}[htb]\centering
\includegraphics*[height=5.2cm,width= 5.5cm]{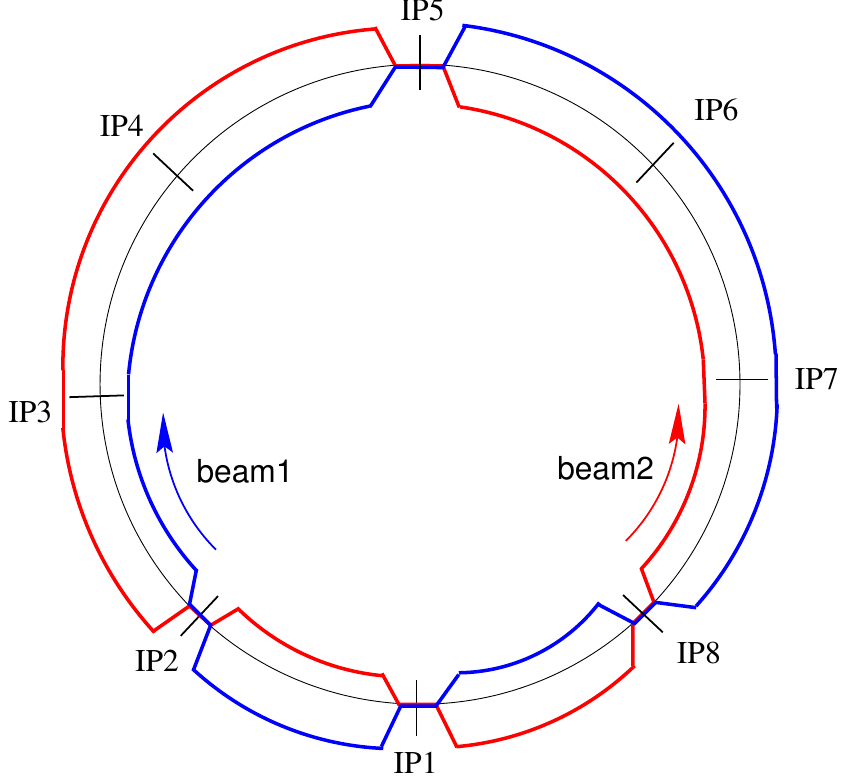}
\caption{Schematic layout of the LHC collision points and beams}\label{fig:011}
\end{figure}

\begin{figure}[htb]\centering
\includegraphics*[height=4.2cm,width= 8.5cm]{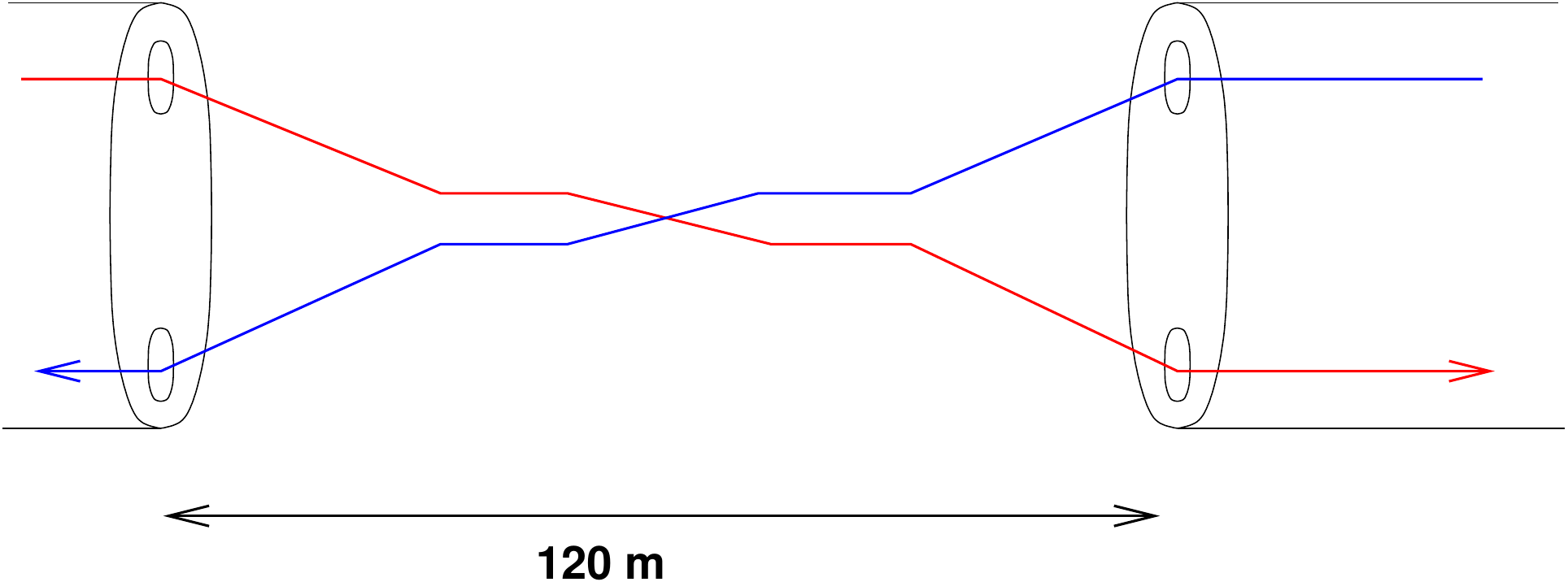}
\caption{Crossover between inner and outer vacuum chambers in the LHC (schematic)}\label{fig:08}
\end{figure}

At the collision points of the LHC the two beams are brought together and
into collision (Fig.~\ref{fig:011}).
An arrangement of separation and recombination magnets is used for the
purpose of making the beams cross (Fig.~\ref{fig:08}).

During that process it is unavoidable that the beams travel in a common
vacuum chamber for more than 120~m.
In the LHC the time between the bunches is only 25~ns and therefore
the bunches will meet in this region.
In order to avoid the collisions, the bunches collide at a small
crossing angle of 285~$\mu$rad.

\begin{figure}[htb]\centering
\includegraphics*[height=3.2cm,width= 8.5cm]{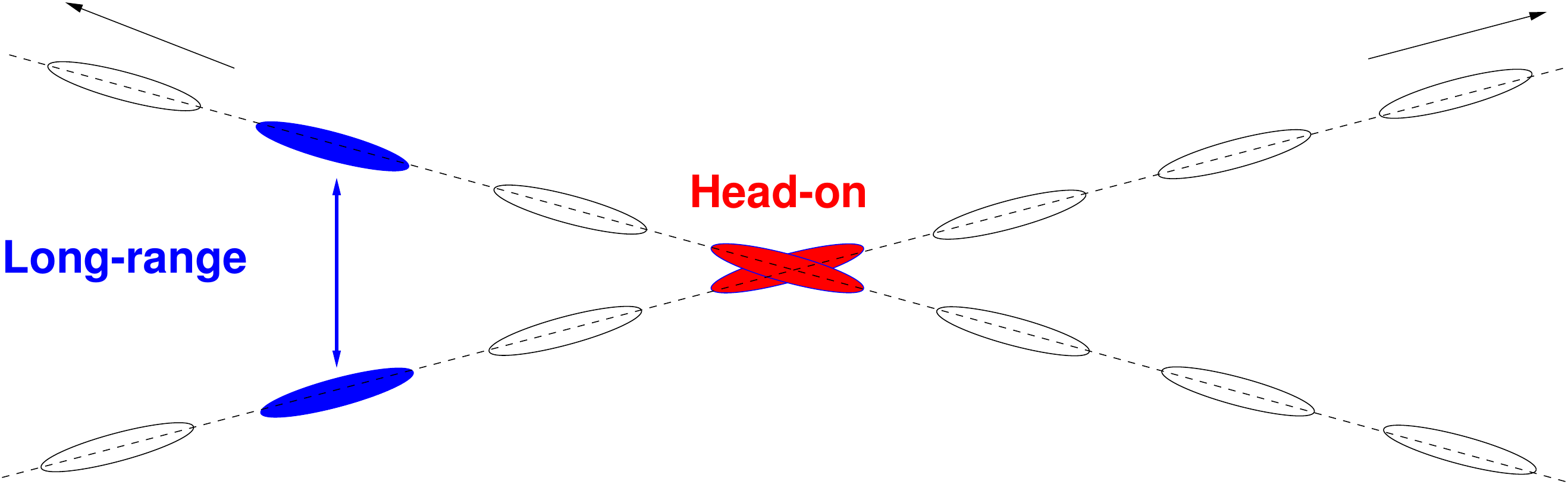}
\caption{Head-on and long-range interactions in a LHC interaction point}\label{fig:07}
\end{figure}

The basic principle is shown in Fig.~\ref{fig:07}: while two bunches collide
at a small angle (quasi head on) at the centre, the other bunches are kept separated
by the crossing angle.
However, since they travel in a common beam pipe, the bunches still feel
the electromagnetic forces from the bunches of the opposite beam.
When the separation is large enough, these so-called long-range interactions
should be weak.
From the bunch spacing and the length of the interaction region one can
easily calculate that at each of the four LHC interaction points we must
expect 30 of these long-range encounters, i.e. in total 120 interactions.
The typical separation between the two beams is between 7 and 10 in units
of the beam size of the opposing beam.

\subsection{Long-range beam--beam effects}
Although the long-range interactions distort the beams much less than
a head-on interaction, their large number and some particular
properties require careful studies.
\begin{itemize}
\item[$\bullet$]They break the symmetry between planes, i.e. odd resonances are also excited.
\item[$\bullet$]While the effect of head-on collisions is strongest for small-amplitude particles, they mostly affect particles at large amplitudes.
\item[$\bullet$]The tune shift caused by long-range interactions has {\em opposite} sign in the plane of separation compared to the head-on tune shift.
\item[$\bullet$]They cause changes of the closed orbit.
\item[$\bullet$]They largely enhance the so-called PACMAN effects.
\end{itemize}

\subsubsection{Opposite-sign tune shift}
The opposite sign of the tune shift can easily be understood when we come back
to the method for calculating the tune spread, explained with the
help of Fig.~\ref{fig:02a}.

\begin{figure}[htb]\centering
\includegraphics*[height=7.2cm,width=14.5cm]{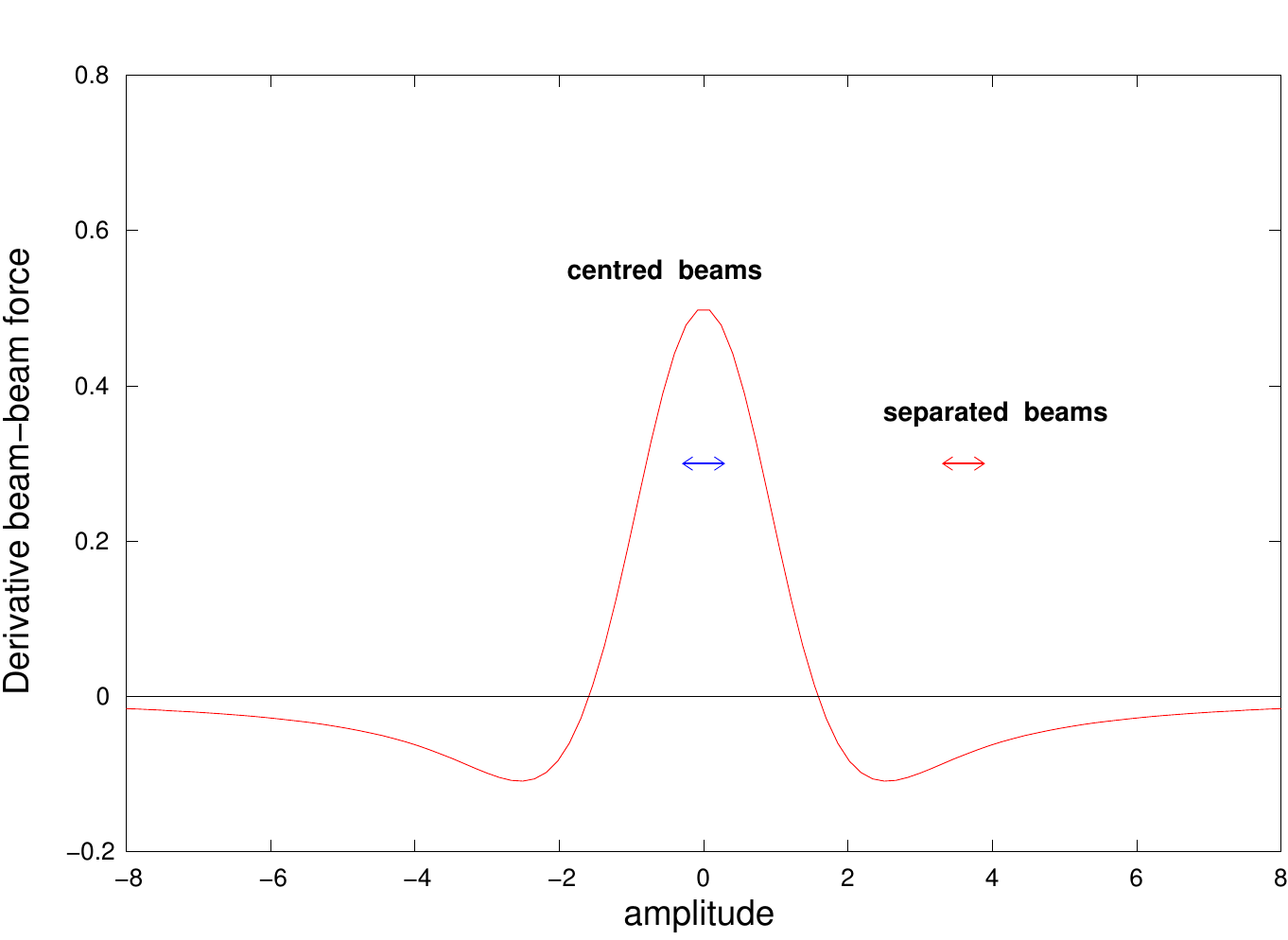}
\caption{Derivative of beam--beam force for round beams. Oscillation range of particles of centred and separated beams.}\label{fig:02c}
\end{figure}

We average again the oscillation of a small-amplitude particle as it samples
the focusing force of the beam--beam interaction.
In Fig.~\ref{fig:02c}, we show the range of oscillation
for central collisions and for the interaction of separated beams, in both cases for particles with small oscillation amplitudes.
When the separation is larger than $\approx1.5\sigma$, the focusing
(slope of the force as a function of the amplitude) changes
sign and the resulting tune shift assumes the opposite sign.

To some extent, this property could be used to partially compensate
long-range interactions when a configuration is used where the
beams are separated in the horizontal plane in one interaction
region and in the vertical plane in another one.

\subsubsection{Strength of long-range interactions}
\begin{figure}[htb]\centering
\includegraphics*[height=4.6cm,width=12.5cm]{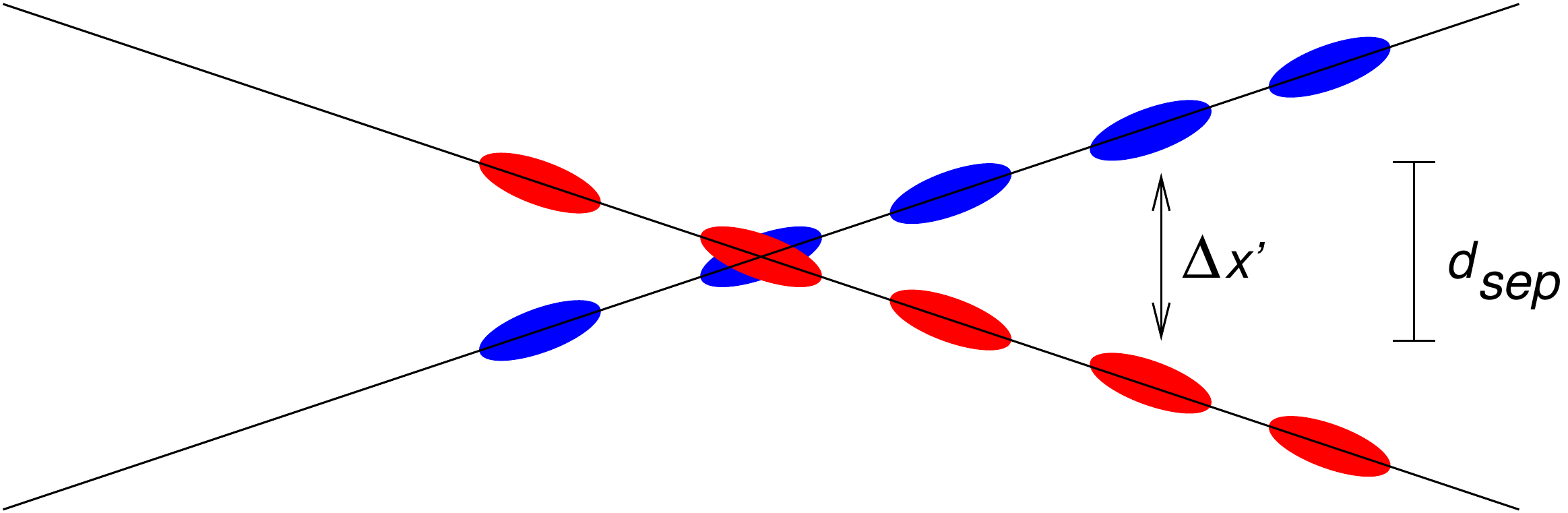}
\caption{Long-range interaction, schematic}\label{fig:010}
\end{figure}

The geometry of a single encounter is shown in Fig.~\ref{fig:010}.
The particles in the test bunch receive a kick (change of
slope) $\Delta x'$.

Assuming a separation $d$ in the horizontal plane,
the kicks in the two planes can be written as
\begin{equation}\label{eq:021}
\Delta x' = -\frac{2 N r_{0}}{\gamma}
\cdot \frac{(x + d)}{r^{2}}\cdot\left[ 1 - {\mathrm{exp}}\left(-\frac{{{r^{2}}}}{2\sigma^{2}}\right)\right]
\end{equation}
with $r^{2} = (x + d)^{2} + y^{2}$.
The equivalent formula for the plane orthogonal to the separation is
\begin{equation}\label{eq:022}
\Delta y' = -\frac{2 N r_{0}}{\gamma}
\cdot\frac{y}{r^{2}}\cdot\left[ 1 - {\mathrm{exp}}\left(-\frac{{{r^{2}}}}{2\sigma^{2}}\right)\right].
\end{equation}
It is fairly obvious that the effect of long-range interactions must strongly
depend on the separation.
The calculation shows that the tune spread $\Delta Q_{\rm lr}$
from long-range interactions alone follows an approximate
scaling (for large enough separation, i.e. above $\approx6\sigma$):
\begin{equation}\label{eq:023}
      \Delta Q_{\rm lr}  \propto  -\frac{N}{d^{2}},
\end{equation}
where $N$ is the bunch intensity and $d$ the separation.
Small changes in the separation can therefore result in significant differences.

\subsubsection{Footprint for long-range interactions}
Contrary to the head-on interaction where the small-amplitude particles
are mostly affected, now the large-amplitude particles experience the
strongest long-range beam--beam perturbations.
This is rather intuitive, since the large-amplitude particles are
the ones which can come closest to the opposing beam as they perform
their oscillations.
We must therefore expect a totally different tune footprint.

\begin{figure}[htb]\centering
\includegraphics*[height=7.2cm,width=12.5cm]{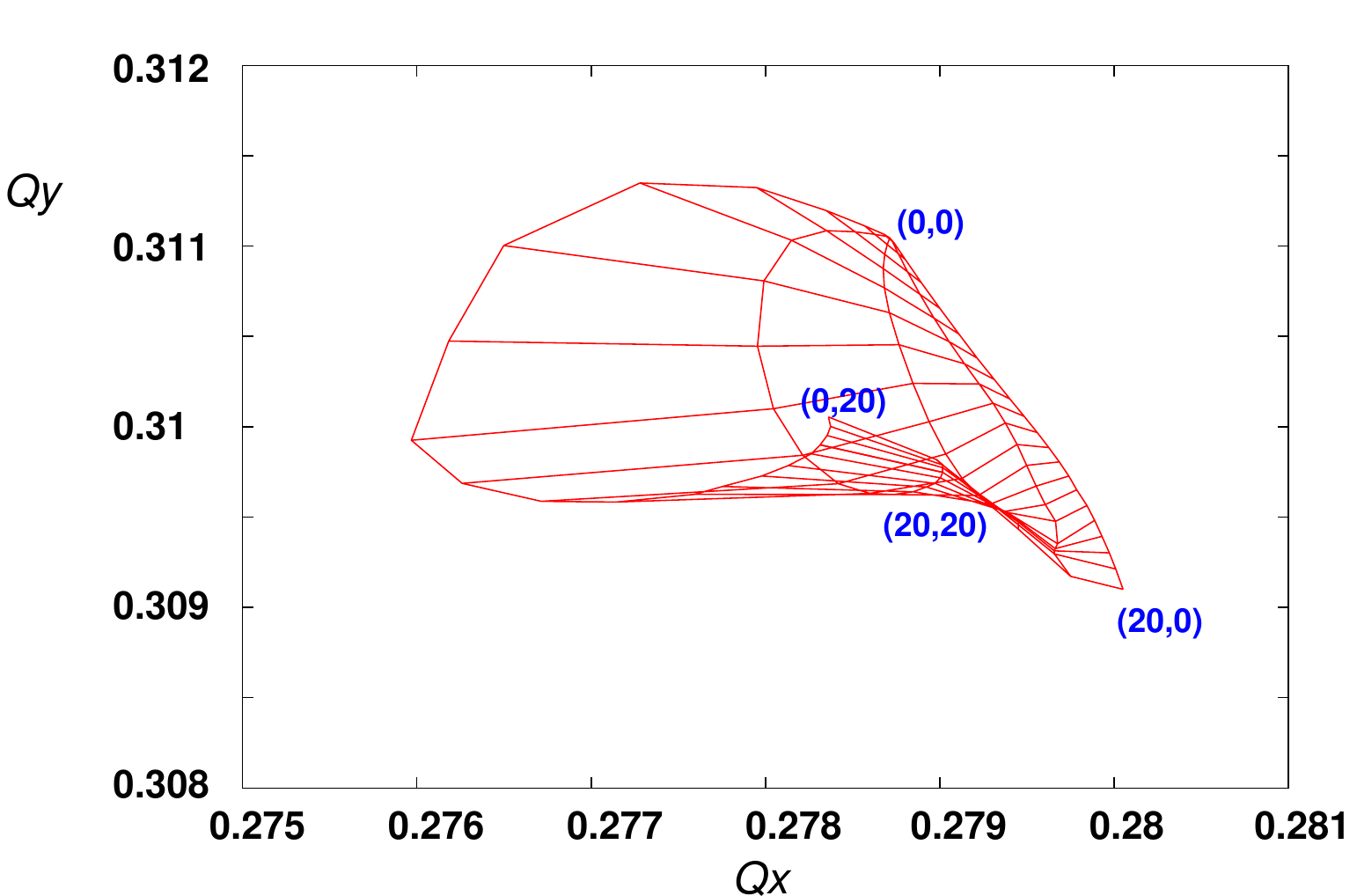}
\caption{Tune footprint for long-range interactions only. Vertical separation and amplitudes between 0 and 20$\sigma$}\label{fig:014}
\end{figure}

Such a footprint for only long-range interactions is shown in
Fig.~\ref{fig:014}.
Since the symmetry between the two planes is broken, the resulting
footprint shows no symmetry.
In fact, the tune shifts have different signs for $x$ and $y$, as
expected.
For large amplitude one observes a `folding' of the footprint.
This occurs when large-amplitude particles are considered,
for which the oscillation
amplitudes extend across the central maximum in Fig.~\ref{fig:02c},
i.e. when the oscillation amplitude is larger than the separation
between the beams.

Such large amplitudes are treated in Fig.~\ref{fig:02c} to demonstrate
this feature.
In practice, these amplitudes are usually not important since
in real machines no particles reach these amplitudes.

\begin{figure}[htb]\centering
\includegraphics*[height=7.2cm,width= 7.2cm]{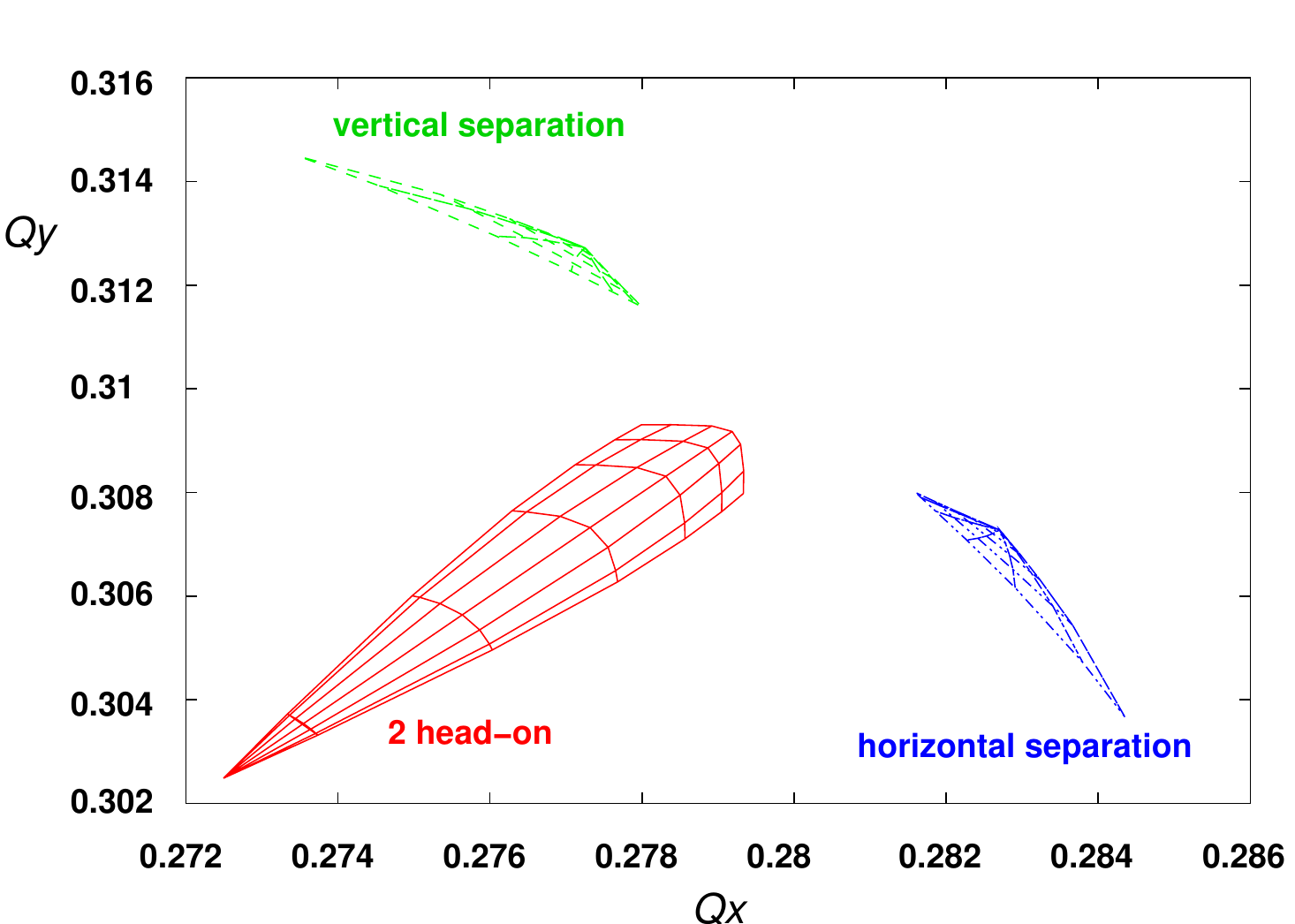}
\includegraphics*[height=7.2cm,width= 7.2cm]{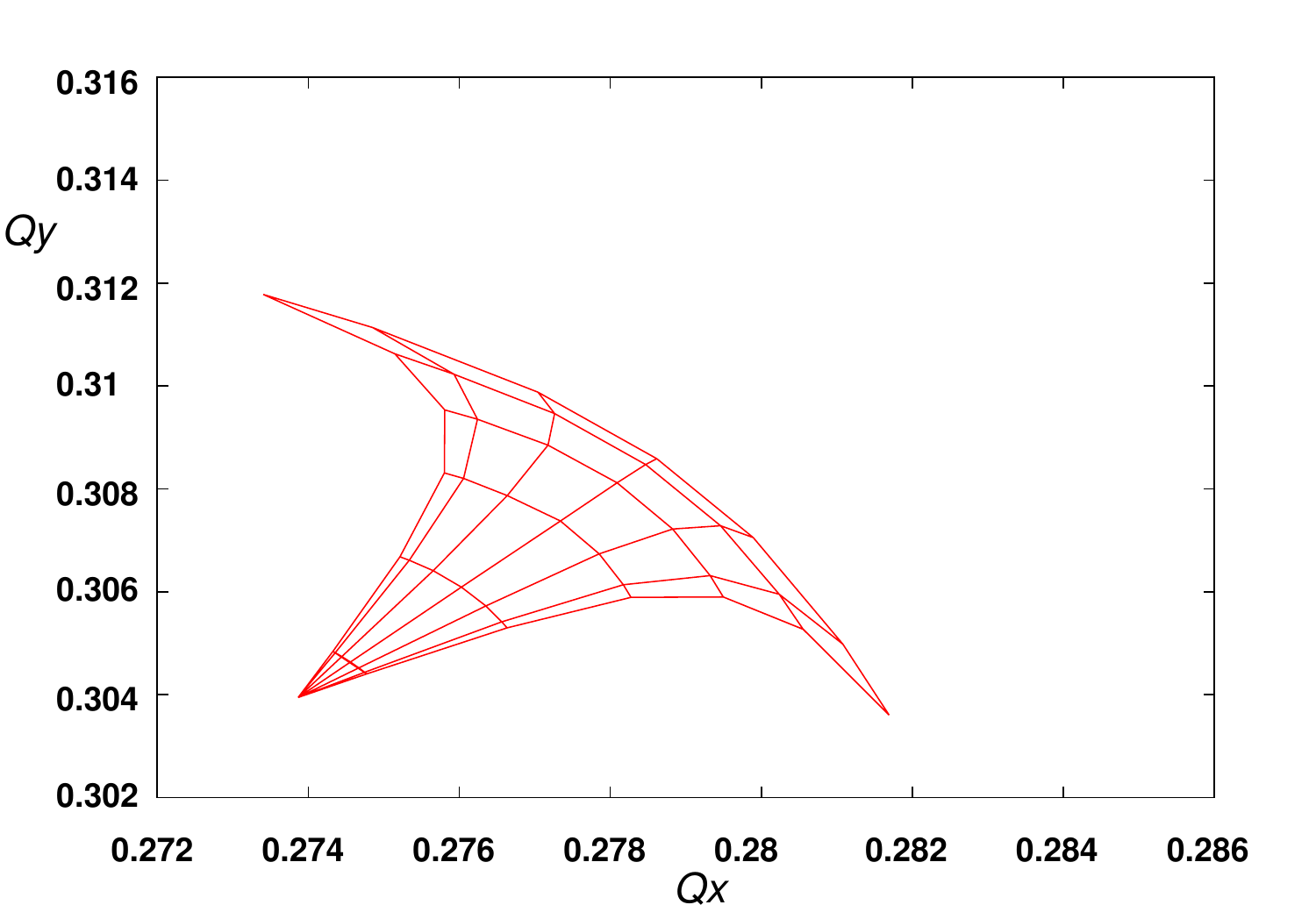}
\caption{Footprint for two head-on interactions (proton--proton), long-range interactions in the horizontal plane and long-range interactions in the vertical plane (left). Combined head-on and long-range interactions, one horizontal and one vertical crossing (right).}\label{fig:015}
\end{figure}

In Fig.~\ref{fig:015} (left), we show separately the footprints for two head-on
proton--proton collisions, and long-range interactions with
horizontal separation and vertical separation, respectively.

The differences and in particular the different sign of the long-range tune shift
are clearly visible.
The long-range footprints are shifted away from the original
tune (0.28, 0.31) in opposite directions.
Amplitudes between 0 and 6$\sigma$ are shown in the figure.
In the same figure on the right we show the combined footprint, i.e. for
particles which experience two head-on collisions, long-range interactions
in one interaction point with horizontal separation and a second with
vertical separation.
A partial compensation can be seen and the footprint is again
symmetric in $x$ and $y$.
In particular, the linear tune shifts of the central parts are very well
compensated.

\subsubsection{Dynamic aperture reduction due to long-range interactions}
For too small separation, the tune spread induced by long-range interactions
can become very large and resonances cannot be avoided any more.
The motion can become irregular and as a result particles at large
amplitudes can get lost.

\begin{figure}[htb]\centering
\includegraphics*[height=5.2cm,width= 6.5cm]{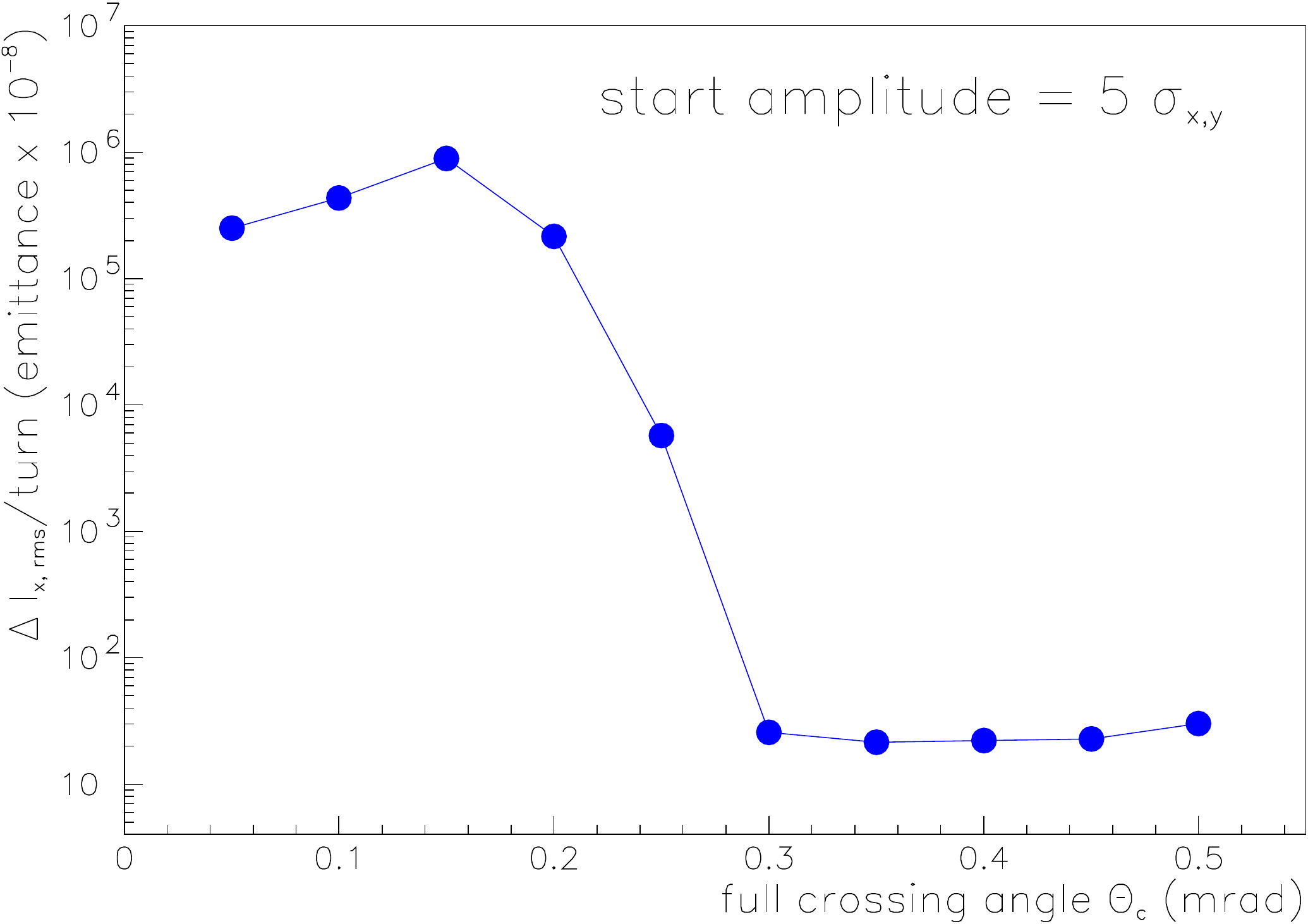}
\caption{Emittance increase as function of crossing angle}\label{fig:017}
\end{figure}

This is demonstrated in Fig.~\ref{fig:017}.
The emittance increase of a large-amplitude particle (5$\sigma$) is
computed for different values of the crossing angle and hence of
the beam separation.
For large enough angles the emittance increase is very small, but it
increases over several orders of magnitude when the crossing angle
drops significantly below 300~$\mu$rad.
These results are obtained by particle tracking \cite{dynap1, dynap2}.

To evaluate the dynamic aperture in the presence of beam--beam interactions,
a simulation of the complete machine is necessary and the interplay
between the beam--beam perturbation and possible machine imperfections
is important \cite{dynap3}.

For the present LHC parameters we consider the minimum crossing angle
to be 285 $\mu$rad.

\section{Beam--beam-induced orbit effects}
\begin{figure}[htb]\centering
\includegraphics*[height=4.5cm,width= 9.0cm]{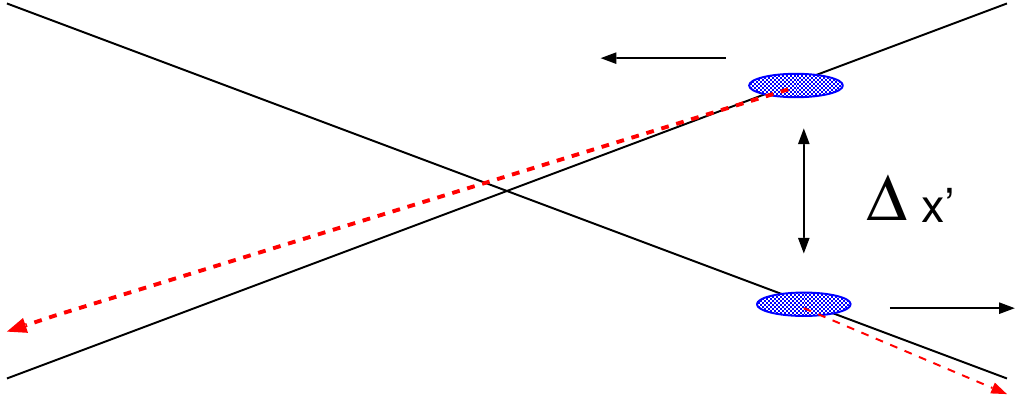}
\caption{Beam--beam deflection leading to orbit changes}\label{fig:018}
\end{figure}

When two beams do not collide exactly head on,
the force has a constant contribution, which can easily be seen when
the kick $\Delta x'$ (from (\ref{eq:021}), for sufficiently large separation)
is developed in a series:
\begin{equation}\label{eq:024}
\Delta x'~~=~~\frac{\mathrm{const}.}{d}~\cdot~\left[~~1~~-~~\frac{x}{d}~~+~~O\left(\frac{x^{2}}{d^{2}}\right)~~+~~\cdots \right].
\end{equation}
A constant contribution, i.e. more precisely an amplitude-independent contribution,
changes the orbit of the bunch as a whole (Fig.~\ref{fig:018}).

When the beam--beam effect is strong enough, i.e. for high intensity and/or
small separation, the orbit effects are large enough to be observed.

When the orbit of a beam changes, the separation between the beams will
change as well, which in turn will lead to a slightly different beam--beam
effect and so on.
The orbit effects must therefore be computed in a self-consistent
way \cite{aaa65}, in particular when the effects are sizable.
The closed orbit of an accelerator can usually be corrected; however,
an additional effect, which is present in some form in many
colliders, sets a limit to the correction possibilities.
A particularly important example is the LHC and we shall therefore
use it to illustrate this feature.

\section{PACMAN bunches}
\begin{figure}[htb]\centering
\includegraphics*[height=7.0cm,width= 9.0cm]{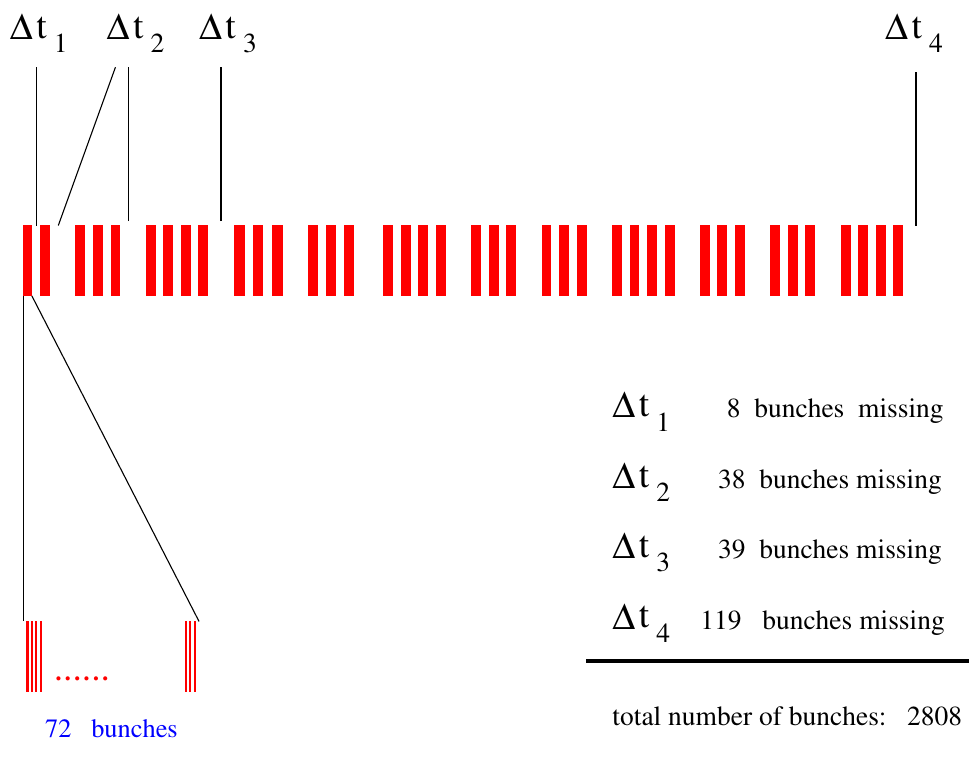}
\caption{Bunch filling scheme in the LHC}\label{fig:019}
\end{figure}

The bunches in the LHC do not form a continuous train of
equidistant bunches spaced by 25~ns, but some empty space must
be provided to allow for the rise time of kickers.
These gaps and the number of bunches per batch are determined by
requirements from the LHC injectors (PS, SPS etc) and the preparation
of the LHC beam (bunch splitting).
The whole LHC bunch pattern is composed of 39 smaller batches (trains
of 72 bunches) separated by gaps of various length
followed by a large abort gap for the dump kicker at
the end.
Figure~\ref{fig:019} shows the
actual LHC filling scheme with the various gaps in the train.

In the LHC, only 2808 out of 3564 possible bunches are present
with the above filling scheme.
Due to the symmetry, bunches normally meet other bunches at the
head-on collision point.
For the long-range interactions this is no longer the case.

\begin{figure}[htb]\centering
\includegraphics*[height=4.5cm,width= 9.0cm]{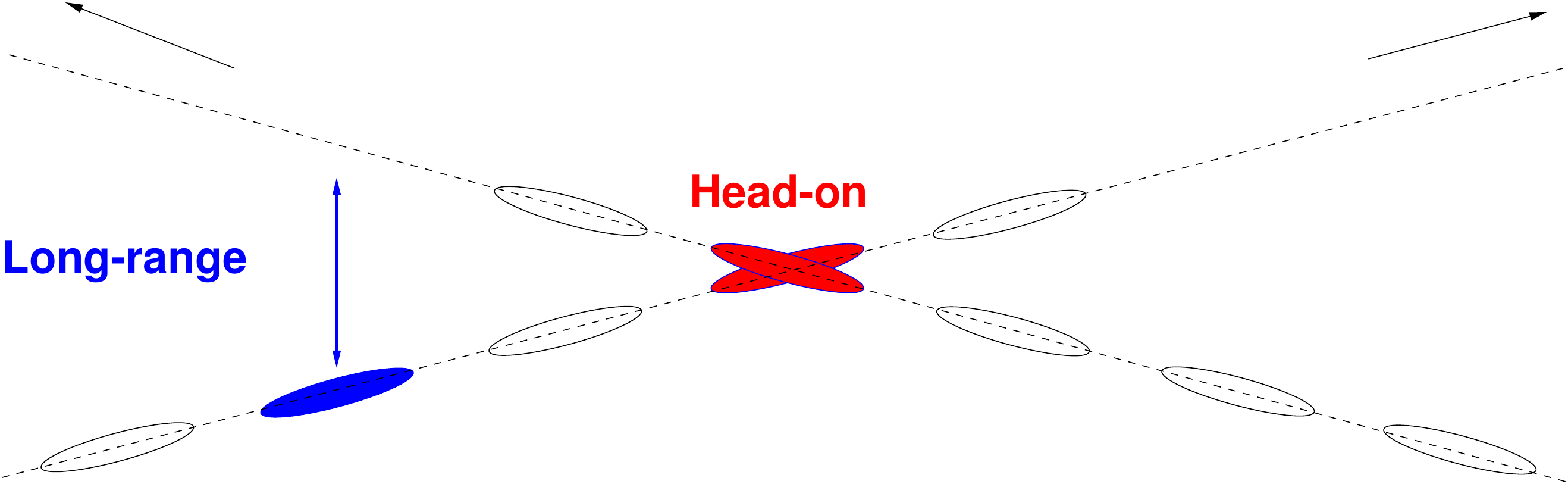}
\caption{Long-range interactions with missing bunches}\label{fig:020}
\end{figure}

This is illustrated in Fig.~\ref{fig:020}. Bunches at the beginning
and at the end of a small batch will encounter a hole and as a result
experience fewer long-range interactions than bunches from the
middle of a batch \cite{aaaa7}.
In the limit, the first bunch of a batch near a large gap encounters no
opposing bunch before the central collision and the full
number of bunches after.

Bunches with fewer long-range interactions have a very different
integrated beam--beam effect and a
different dynamics must be expected.
In particular, they will have a different tune and occupy a different
area in the working diagram; therefore, they may be susceptible to
resonances which can be avoided for nominal bunches.
The overall space needed in the working diagram
is therefore largely increased \cite{aaaa7, cross}.

Another consequence of reduced long-range interactions is the
different effect on the closed orbit of the bunches.
We have to expect a slightly different orbit from bunch to bunch.

\begin{figure}[htb]\centering
\includegraphics*[height=7.0cm,width=11.5cm]{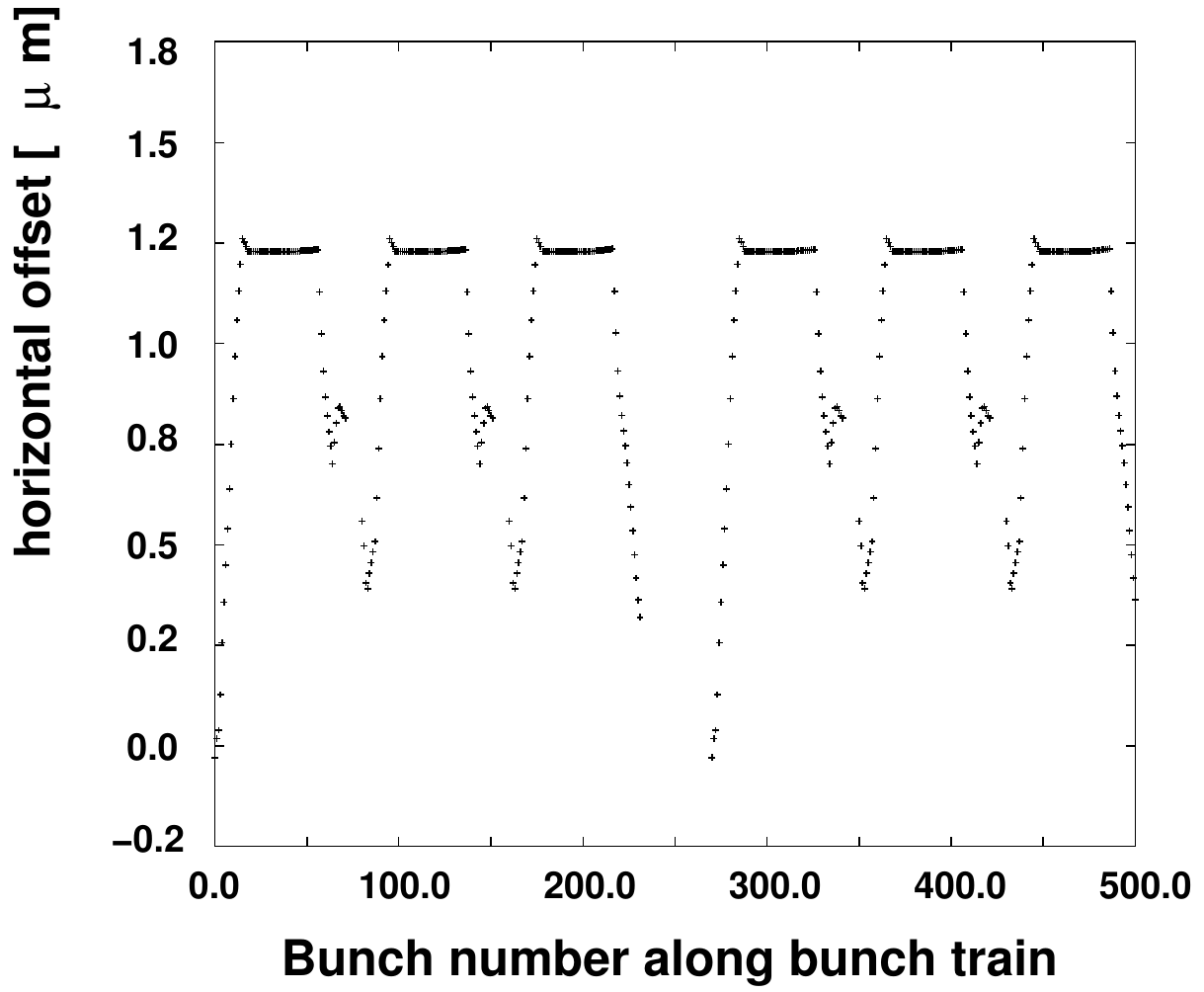}
\caption{Horizontal orbits of the first 432 bunches at IP1}\label{fig:021}
\end{figure}

This effect is demonstrated in Fig.~\ref{fig:021}, where we show the
horizontal position at one head-on collision point for 432 bunches
(out of 2808).
The bunches in the middle of a batch have all interactions and
therefore the same orbit while the bunches at the beginning and
end of a batch show a structure which exhibits the decreasing
number of long-range interactions.
The orbit spread is approximately 10--15\% of the beam size.
Since the orbits of the two beams are not the same, it is impossible
to make all bunches collide exactly head on.
A significant fraction will collide with an offset.
Although the immediate effect on the luminosity
is small \cite{aaa63}, collisions at an offset can potentially
affect the dynamics and are undesirable.
The LHC design should try to minimize these offsets \cite{aaaa7, cross}.

\begin{figure}[htb]\centering
\includegraphics*[height=7.0cm,width=11.5cm]{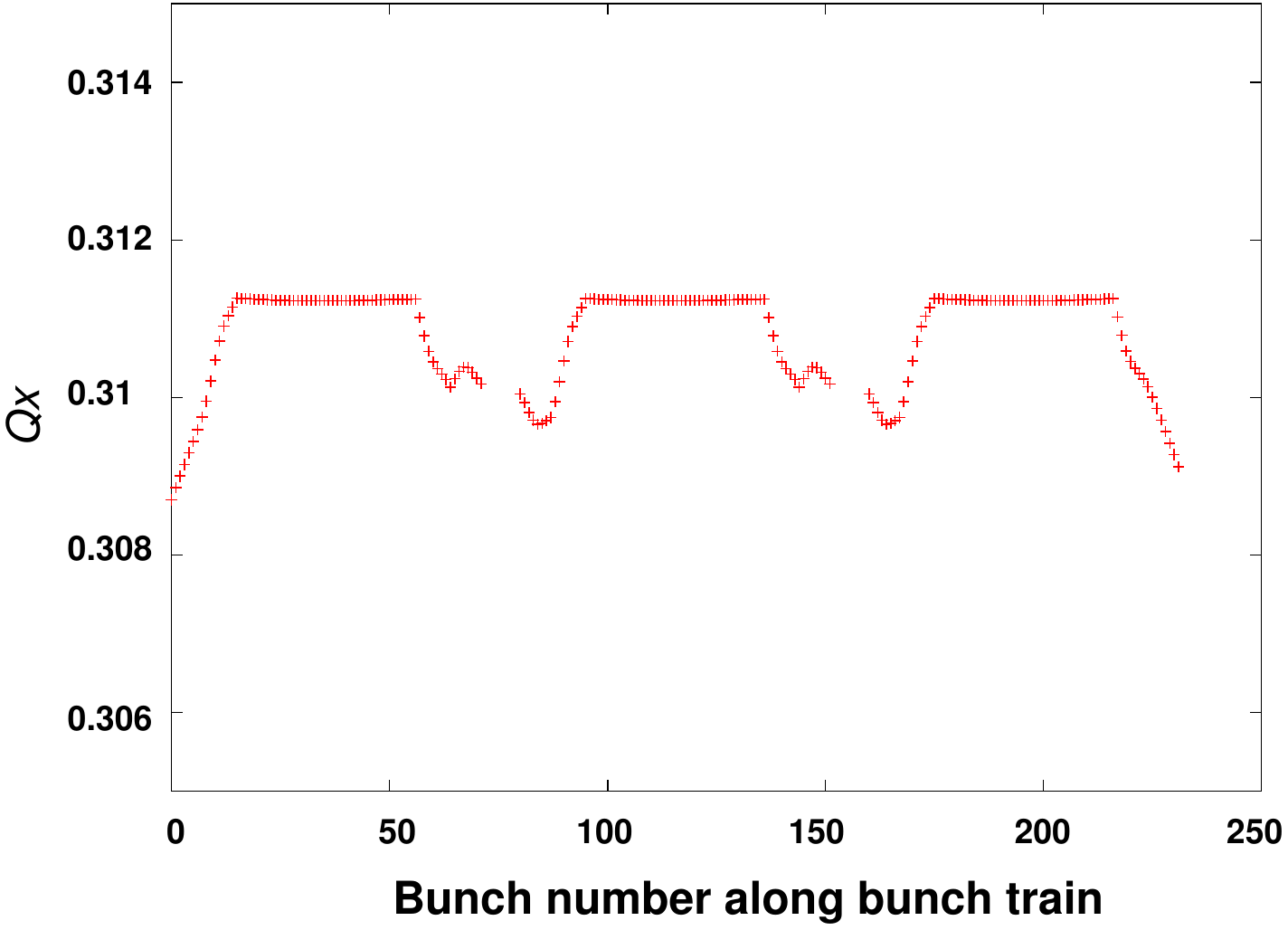}
\caption{Tune variation along the LHC bunch pattern}\label{fig:022}
\end{figure}

A second effect, the different tunes of the bunches, is shown in
Fig.~\ref{fig:022}.
For three batches it shows a sizable spread from bunch to bunch and
without compensation effects \cite{cross} it may be too large for
a safe operation.

\section{Coherent beam--beam effects}
So far we have mainly studied how the beam--beam interaction affects the single-particle behaviour and treated the beam--beam interaction as
a static lens.
In the literature this is often called a `weak--strong' model:
a `weak' beam (a single particle) is perturbed by a `strong'
beam (not affected by the weak beam).
When the beam--beam perturbation is important, the model of an
unperturbed, strong beam is not valid any more, since its
parameters change under the influence of the other beam and
vice versa.
When this is the case, we talk about so-called `strong--strong' conditions.
The first example of such a `strong--strong' situation
was the orbit effect where the beams mutually changed their
closed orbits.
These closed orbits had to be found in a self-consistent way.
This represents a static strong--strong effect.

In the next step we investigate dynamic effects under the
strong--strong condition \cite{aaa12}. When we consider the coherent motion of bunches, the
collective behaviour of all particles in a bunch is studied.
A coherent motion requires an organized
behaviour of all particles in a bunch.
A typical example are oscillations of the centre of mass of
the bunches, so-called dipole oscillations.
Such oscillations can be driven by external forces such
as impedances and may be unstable.
At the collision of two counter-rotating bunches not only the
individual particles receive a kick from the opposing beam,
but the bunch as an entity gets a coherent kick.
This coherent kick of separated beams can excite coherent dipole oscillations.
Its strength depends on the distance between the bunch centres at the
collision point.
It can be computed by adding the individual contributions of all particles.
For small distances it can be shown \cite{aaa64,hira} that
it is just one-half of the incoherent kick a single particle
would receive at the same distance.
For distances large enough, the incoherent and coherent kicks become
the same.

\section{Coherent beam--beam modes}
To understand the dynamics of dipole oscillations, we first
study the simplest case with one bunch in each beam.

\begin{figure}[htb]
\centering{
\includegraphics*[width= 6.5cm,clip=]{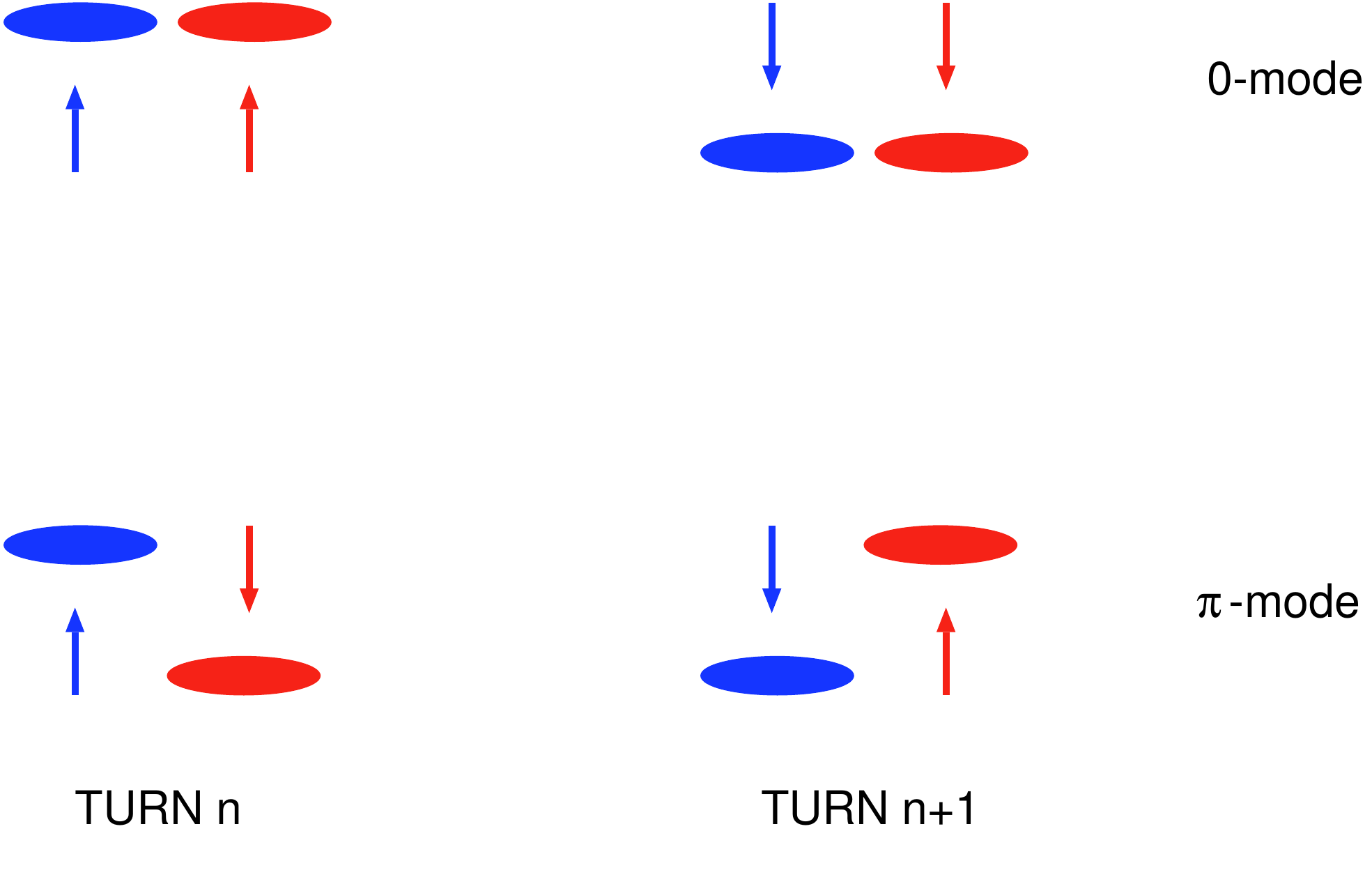}
}
\caption{Basic dipole modes of two bunches. Relative position of the bunches at the interaction point at two consecutive turns.}\label{fig:025}
\label{fig:sigpi}
\end{figure}

When the bunches meet turn after turn
at the collision point, their oscillation
can either be exactly in phase (0 degree phase difference)
or out of phase (180 degrees or $\pi$ phase difference).
Any other oscillation can be constructed from these basic modes.
The modes are sketched very schematically in Fig.~\ref{fig:025}.
The relative positions of the bunches as observed at the interaction
point are shown for two consecutive turns $n$ and $n+1$.
The first mode is called the 0-mode (or sometimes called the $\sigma$-mode)
and the second the $\pi$-mode.
In the first mode the distance between the bunches does not
change turn by turn and therefore there is no net force driving
an oscillation.
This mode must oscillate with the unperturbed frequency (tune) $Q_{0}$.
For the second mode the net force difference between two turns
is a maximum and the tune becomes $Q_{0}+\Delta Q_{\rm coh}$.
The sign of $\Delta Q_{\rm coh}$ depends whether the two beams have
equal charge (defocusing case) or opposite charge (focusing case).
The calculation of $\Delta Q_{\rm coh}$ is non-trivial: when the bunches
are considered as rigid objects, the tune shift can be computed
easily using the coherent kick but is underestimated \cite{piw}.
The correct calculation must allow for changes of the density
distribution during the collision and, moreover, must allow
a deviation from a Gaussian function.
The computation requires us to solve the Vlasov equation of
two coupled beams \cite{mell, aaaa9, aaa10, aaa11}.

The 0-mode is found at the unperturbed tune, as expected.
The $\pi$-mode is shifted by 1.2--1.3$\xi$. The precise value
depends on the ratio of the horizontal and vertical beam sizes \cite{aaaa9}.
We have seen before the incoherent tune spread (footprint) the individual
particles occupy and we know that it spans the interval
[0.0, 1.0] $\xi$, starting at the 0-mode.

Here one can make an important observation: under the strong--strong
condition the $\pi$-mode is a discrete mode outside the incoherent
spectrum \cite{aaa10, aaa11}.
This has dramatic consequences for the stability of the beams.
A coherent mode that is outside an incoherent frequency spectrum
cannot be stabilized by Landau damping.
Under these conditions the coherent beam--beam effect could
drive the dipole oscillation to large amplitudes and may result in
the loss of the beam.
Observations of the coherent beam--beam effects have been made at
PETRA \cite{piw}.
Beam--beam modes have been observed with high-intensity coasting
beams in the ISR \cite{jpk} and
recently in a bunched hadron collider at RHIC \cite{rhic}.

Coherent beam--beam modes can be driven by head-on collisions with
a small offset or by long-range interactions.
In the first case and for small oscillations, the problem
can be linearized and the theoretical treatment is simplified.
The forces from long-range interactions are very non-linear but
the numerical evaluation is feasible.
Since the coherent shift must have the opposite sign for long-range
interactions, the situation is very different.
In particular, the $\pi$-mode from long-range interactions alone
would appear on the opposite side of the 0-mode in the frequency
spectrum \cite{aaa11, aaa14}.
Both the incoherent and the coherent spectra include both types of
interactions.

\section{Compensation of beam--beam effects}
For the case where the beam--beam effects limit the performance of a collider, several schemes have be proposed to compensate
all or part of the detrimental effects.
The basic principle is to design correction devices which
act as non-linear `lenses' to counteract the
distortions from the non-linear beam--beam `lens'.

For both head-on and long-range effects schemes have been
proposed:
\begin{itemize}
\item[$\bullet$]head-on effects:           \\
\begin{itemize}
\item[-]electron lenses;
\item[-]linear lens to shift tunes;
\item[-]non-linear lens to decrease tune spread;     \\
\end{itemize}
\item[$\bullet$]long-range effects:         \\
\begin{itemize}
\item[-]at large distance: beam--beam force changes like $1/r$;
\item[-]same force as a wire.
\end{itemize}
\end{itemize}

\section{Electron lenses}
The basic principle of a compensation of proton--proton (or --antiproton)
collisions with an `electron lens' implies
that the proton (antiproton) beam travels through a counter-rotating high-current
electron beam (`electron lens') \cite{aaa17a, aaa17b}.
The negative electron space charge can reduce the effect from the
collision with the other proton beam.

An electron beam with a size much larger than the proton beam can
be used to shift the tune of the proton beam (`linear lens').
When the current in the electron bunches can be varied fast enough,
the tune shift can be different for the different proton bunches,
thus correcting PACMAN tune shifts.

When the electron charge distribution is chosen to be the same
as the counter-rotating proton beam, the non-linear focusing of
this proton beam can be compensated (`non-linear lens').
When it is correctly applied, the tune spread in the beam can be
strongly reduced.

Such lenses have been constructed at the Tevatron at
Fermilab \cite{aaa17b} and experiments are in progress.

\section{Electrostatic wire}
To compensate the tune spread from long-range interactions,
one needs a non-linear lens that resembles a separated beam.
At large enough separation, the long-range force changes approximately
with $\frac{1}{r}$ and this can be simulated by a wire parallel
to the beam \cite{jpk2}.

In order to compensate PACMAN effects, the wires have to be pulsed
according to the bunch filling scheme.
Tests are in progress at the SPS to study the feasibility of
such a compensation for the LHC.

\section{M\"{o}bius scheme}
The beam profiles of e$^{+}$e$^{-}$ colliders are usually flat,
i.e. the vertical beam size is much smaller than the horizontal
beam size.
Some studies indicate that the collision of round beams, even
for e$^{+}$e$^{-}$ colliders, shows more promise for higher
luminosity since larger beam--beam parameters can be achieved.
Round beams can always be produced by strong coupling between
horizontal and vertical planes.
A more elegant way is the so-called M\"{o}bius
lattice \cite{aaa18, aaa19}.
In this lattice, the horizontal and vertical betatron oscillations
are exchanged by an insertion.
A horizontal oscillation in one turn becomes a vertical oscillation
in the next turn and vice versa.
\
Tests with such a scheme have been done at CESR at Cornell \cite{aaa19}.

\appendix
\section{Appendix A}
In practice, one usually derives the potential $U(x, y, z)$ from the Poisson equation, which relates the potential to the charge-density distribution $\rho(x, y, z)$:
\begin{equation}\label{eq:00a01}
\Delta U = - \frac{1}{\epsilon_{0}} \rho(x, y, z)
\end{equation}
and computes the fields from
\begin{equation}
\vec{E}~=~-\nabla U(x,y,\sigma_{x},\sigma_{y}).
\end{equation}
The Poisson equation can be solved using e.g. the Green's function
method (e.g. \cite{webs}) since the Green's function for this boundary value
problem is well known.
The formal solution using a Green's function $G(x, y, z, x_{0}, y_{0}, z_{0})$ is
\begin{equation}\label{eq:00a02}
U(x,y,z) = \frac{1}{\epsilon_{0}}\int G(x, y, z, x_{0}, y_{0}, z_{0}) \cdot \rho(x_{0}, y_{0}, z_{0}) {\mathrm d}x_{0} {\mathrm d}y_{0} {\mathrm d}z_{0}.
\end{equation}
For the solution of the Poisson equation, we get \cite{bron}
\begin{equation}\label{eq:00a1}
U(x,y,z,\sigma_{x},\sigma_{y},\sigma_{z}) =
\frac{1}{4 \pi \epsilon_{0}}\int \int \int \frac{\rho(x_{0}, y_{0}, z_{0}) {\mathrm d}x_{0} {\mathrm d}y_{0} {\mathrm d}z_{0}}{\sqrt{(x - x_{0})^{2} + (y - y_{0})^{2} + (z - z_{0})^{2}}}.
\end{equation}
In the case of a beam with Gaussian beam density distributions,
we can factorize the density distribution
$\rho(x_{0},y_{0},z_{0})~=~\rho(x_{0})~\cdot~\rho(y_{0})~\cdot~\rho(z_{0})$
with r.m.s. of $\sigma_{x}$, $\sigma_{y}$ and $\sigma_{z}$:
\begin{equation}
\rho(x_{0},y_{0},z_{0}) = \frac{N e}{\sigma_{x}\sigma_{y}\sigma_{z}({\sqrt{2\pi}})^{3}}
\mathrm{e}^{\left(
-\frac{x_{0}^{2}}{2\sigma_{x}^{2}}
-\frac{y_{0}^{2}}{2\sigma_{y}^{2}}
-\frac{z_{0}^{2}}{2\sigma_{z}^{2}}
\right)}.
\end{equation}
Here $N$ is the number of particles in the bunch.
We therefore have
\begin{equation}\label{eq:00a2}
U(x,y,z,\sigma_{x},\sigma_{y},\sigma_{z}) =
\frac{1}{4 \pi \epsilon_{0}}\frac{N e}{\sigma_{x}\sigma_{y}\sigma_{z}({\sqrt{2\pi}})^{3}}
\int \int \int
\frac{\mathrm{e}^{\left(
-\frac{x_{0}^{2}}{2\sigma_{x}^{2}}
-\frac{y_{0}^{2}}{2\sigma_{y}^{2}}
-\frac{z_{0}^{2}}{2\sigma_{z}^{2}}
\right)}
{\mathrm d}x_{0} {\mathrm d}y_{0} {\mathrm d}z_{0}}
{{\sqrt{(x - x_{0})^{2} + (y - y_{0})^{2} + (z - z_{0})^{2}}}}.
\end{equation}
This is difficult to solve and we rather follow the proposal by~Kheifets~\cite{aaa52}
and solve the diffusion equation
\begin{equation}\label{eq:00a3}
\Delta V  -  A^{2}\cdot\frac{\delta V}{\delta t}= -\frac{1}{\epsilon_{0}}\rho(x, y, z)~~~~~~({\mathrm{for}~~t}\geq 0)
\end{equation}
and obtain the potential $U$ by going to the limit $A~\rightarrow~0$, i.e.
\begin{equation}\label{eq:00a4}
U  = \lim_{A \rightarrow 0}  V.
\end{equation}
The reason for this manipulation is that the Green's function to solve the
diffusion equation takes a more appropriate form \cite{bron}:
\begin{equation}\label{eq:00a5}
G(x, y, z, t, x_{0}, y_{0}, z_{0}) =
\frac{A^{3}}{(2\sqrt{\pi t})^3} \cdot
\mathrm{e}^{-A^{2}/4t\cdot((x - x_{0})^{2} + (y - y_{0})^{2} + (z - z_{0})^{2})}
\end{equation}
and we get for $V(x,y,z,\sigma_{x},\sigma_{y},\sigma_{z})$:
\begin{equation}\label{eq:00a51}
\frac{N e}{\sigma_{x}\sigma_{y}\sigma_{z}({\sqrt{2\pi}})^{3}\epsilon_{0}}
\int_{0}^{t} \mathrm{d}\tau \int \int \int
\mathrm{e}^{\left(
-\frac{x_{0}^{2}}{2\sigma_{x}^{2}}
-\frac{y_{0}^{2}}{2\sigma_{y}^{2}}
-\frac{z_{0}^{2}}{2\sigma_{z}^{2}}
\right)}
\frac{A^{3} \cdot \mathrm{e}^{-A^{2}/4\tau((x - x_{0})^{2} + (y - y_{0})^{2} + (z - z_{0})^{2})}}{(2\sqrt{\pi\tau})^{3}}
{\mathrm d}x_{0} {\mathrm d}y_{0} {\mathrm d}z_{0}.
\end{equation}
This allows us to avoid the denominator in the integral and to collect the exponential
expressions, which can then be integrated.
Changing the independent variable $\tau$ to $q~=~{4 \tau}/{A^{2}}$
and using the formula \cite{bron} for the three integrations,
\begin{equation}\label{eq:00a6}
\int_{-\infty}^{\infty} \mathrm{e}^{-(au^{2} + 2bu + c)} {\mathrm d}u ~=~\sqrt{\frac{\pi}{a}} \mathrm{e}^{(\frac{b^{2} - ac}{a})} ~~~~~~(\mathrm{for}~~~u = x_{0}, y_{0}, z_{0}),
\end{equation}
we can integrate (\ref{eq:00a02}) and with (\ref{eq:00a4}) we get the
potential $U(x,y,z,\sigma_{x},\sigma_{y},\sigma_{z})$ \cite{aaa51, aaa52}:
\begin{equation}\label{eq:00a7}
U(x,y,z,\sigma_{x},\sigma_{y},\sigma_{z}) = \frac{1}{4 \pi \epsilon_{0}}\frac{N e}{\sqrt{\pi}}
\int_{0}^{\infty} \frac{{\mathrm{exp}}\left(-\frac{x^{2}}{2\sigma^{2}_{x}+q} -\frac{y^{2}}{2\sigma^{2}_{y}+q} -\frac{z^{2}}{2\sigma^{2}_{z}+q}\right)}{\sqrt{(2\sigma^{2}_{x} + q)(2\sigma^{2}_{y} + q)(2\sigma^{2}_{z} + q)}} {\mathrm d}q.
\end{equation}
Since we are interested in the transverse fields, we can work with
the two-dimensional potential (Appendix~A):
\begin{equation}\label{eq:00a8}
U(x,y,\sigma_{x},\sigma_{y}) = \frac{n e}{4\pi\epsilon_{0}}
\int_{0}^{\infty} \frac{{\mathrm{exp}}\left(-\frac{x^{2}}{2\sigma^{2}_{x}+q} -\frac{y^{2}}{2\sigma^{2}_{y}+q}\right)}{\sqrt{(2\sigma^{2}_{x} + q)(2\sigma^{2}_{y} + q})} {\mathrm d}q,
\end{equation}
where $n$ is the line density of particles in the beam, $e$ the elementary
charge and $\epsilon_{0}$ the electrostatic constant.
In this case we do not yet make any assumptions on the longitudinal
distribution.
\newpage

\section{Appendix B}
For the one-dimensional case, we write the betatron motion of a single
particle as a simple harmonic oscillator and use the `smooth approximation':
\begin{equation}\label{eq:00a9}
\end{equation}
\begin{equation}\label{eq:0a10}
  r = R~\cos (\Phi)
\end{equation}
and its derivative:
\begin{equation}\label{eq:0a11}
  r' = -\frac{R}{\beta}~\sin (\Phi).
\end{equation}

\begin{figure}[htb]
\centering{
\includegraphics*[width= 10.2cm, height=10.2cm]{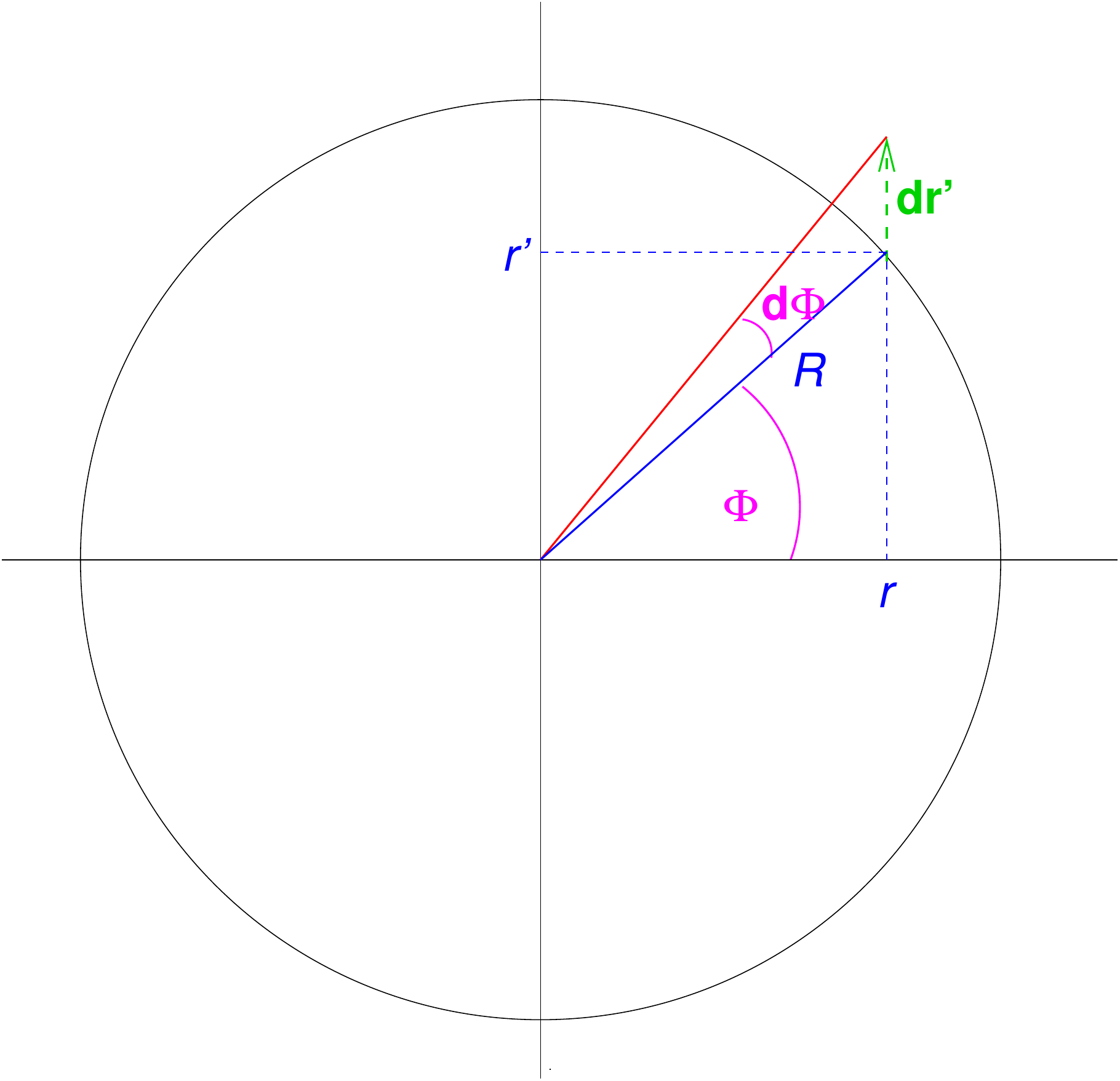}
}
\caption{Phase space before and after the beam--beam kick. Change of phase $\mathrm{d}\Phi$}\label{fig:031}
\label{fig:a01}
\end{figure}

After a small kick from the beam--beam interaction the phase $\Phi$
will be shifted and
we can calculate $\mathrm{d}\Phi$, which is
the instantaneous tune change $\Delta Q_{\rm i}$ times 2$\pi$ (Fig.~\ref{fig:031}).
We have \cite{aaaa2, aaaob}
\begin{equation}\label{eq:0a12}
       2\pi \Delta Q_{\rm i}~=~-\frac{\mathrm{d}r' \cdot \cos(\Phi) \cdot \beta}{R}.
\end{equation}
The deflection $\mathrm{d}r'$ we have calculated in (\ref{eq:013}):
\begin{equation}\label{eq:0a13}
\mathrm{d}r'(r) = -\frac{ 2 N r_{0}}{\gamma r}
\left[ 1 - {\mathrm{exp}}\left(-\frac{{{r^{2}}}}{2\sigma^{2}}\right)\right]
\end{equation}
and linearized for small amplitudes $r$ it becomes (see Eq. (\ref{eq:015}))
\begin{equation}\label{eq:0a14}
  \delta r'|_{r\rightarrow 0}~~=~~-\frac{N r_{0}\cdot r}{\gamma \sigma^{2}}~~=~~-\frac{N r_{0}\cdot R \cos(\Phi)}{\gamma \sigma^{2}},
\end{equation}
\begin{equation}\label{eq:0a15}
  2\pi \Delta Q_{\rm i}~~=~~-\frac{N r_{0}\cdot R \cos(\Phi)}{\gamma \sigma^{2}}~\cdot~\frac{\cos(\Phi) \beta}{R},
\end{equation}
\begin{equation}\label{eq:0a16}
  \Delta Q_{\rm i}~~=~~-\frac{N r_{0} \beta}{2\pi \gamma \sigma^{2}}~\cdot~\cos^{2}(\Phi).
\end{equation}
After averaging $\Phi$ from 0 to 2$\pi$:
\begin{equation}\label{eq:0a17}
  \Delta Q~~=~~\frac{1}{2\pi} \int_{0}^{2\pi} \Delta Q_{\rm i}  {\mathrm d}\Phi
\end{equation}
we get the linear beam--beam tune shift:
\begin{equation}\label{eq:0a18}
  \Delta Q~~=~\xi~~=~~-\frac{N r_{0} \beta}{4\pi \gamma \sigma^{2}}.
\end{equation}

For the non-linear tune shift we must not linearize the beam--beam force and
get for the instantaneous tune shift:
\begin{equation}\label{eq:0a19}
  \Delta Q_{\rm i,nl}~~=~~-\frac{N r_{0} \beta}{\pi \gamma}~\cdot~\frac{1 - \mathrm{e}^{-\frac{R^{2}}{2\sigma^{2}}\cos^{2}(\Phi)}}{R^{2}}.
\end{equation}
To perform the integral, we first substitute the $\cos^{2}(\Phi)$ term in
the exponential by the expression
\begin{equation}\label{eq:0a20}
\cos^{2}(\Phi)~=~\frac{1}{2} (1 + \cos(2\Phi))
\end{equation}
and then perform the integral using
\begin{equation}\label{eq:0a21}
\frac{1}{\pi} \int_{0}^{\pi} \mathrm{e}^{x~\cos(\Phi)} {\mathrm d}\Phi = I_{0}(x),
\end{equation}
where $I_{0}(x)$ is the modified Bessel function,
and get the formula for the non-linear detuning with the amplitude $J$:
\begin{equation}\label{eq:0a22}
\Delta Q(J)~=~\xi~\cdot~\frac{2}{J}~\cdot~( 1 - I_{0}(J/2) \cdot \mathrm{e}^{-J/2}),
\end{equation}
which is $J~=~\epsilon \beta/2\sigma^{2}$ in the usual units.
Here $\epsilon$ is the particle `emittance' and not the beam emittance.

\newpage
\section{Appendix C}
\subsection{Classical approach}
A standard treatment to assess non-linear perturbations is the
$s$-dependent Hamiltonian and perturbation theory:
\begin{eqnarray}
{\cal{H}}~=~ {\cal{H}}_{0} + \delta(s) \epsilon V,
\end{eqnarray}
where ${\cal{H}}_{0}$ is the unperturbed part of the Hamiltonian and
$\epsilon V$ describes the perturbation caused at the position $s$, specified
by the $\delta$ function.
The mathematical treatment is rather involved and in most cases cannot be carried
beyond leading order in the perturbation.
This can easily lead to wrong conclusions, which can still be
found in the literature (e.g. fourth-order resonance cannot be driven by sextupoles)
and one must ask the question whether this is the most appropriate tool to deal
with this problem.
In the case of isolated non-linearities caused by very local beam--beam interactions we
favour a map-based approach as promoted by Dragt \cite{dragt, dragt1} and described
in detail by Forest \cite{ef1}.

\subsection{Map-based approach using Lie transforms and invariant}
In this approach the ring is represented by a finite sequence of maps, which describe the individual
elements.
Possible representations of these maps are Taylor maps and Lie maps \cite{chao}.

In this study we shall use the Lie maps which are always symplectic and
other advantages will become obvious \cite{herr1}.
This technique allows to derive invariants of the motion in a straightforward
way; in particular, the extension of the results to multiple beam--beam
interactions becomes an easy task.

To answer the initial question, this is particularly relevant since we want to investigate
the effect of the number of interaction points and the relative phase advance
on the beam dynamics.

In the first part we derive the formulae for a single interaction point and later
extend the method to multiple beam--beam interactions.

\subsubsection{Single interaction point}
The derivation for a single interaction point can be found in the
literature (see a particularly nice derivation by Chao \cite{chao}).

In this simplest case of one beam--beam interaction we can factorize the machine into a
linear transfer map $e^{:f_{2}:}$ and the beam--beam interaction $e^{:F:}$, i.e.
\begin{eqnarray}
e^{:f_{2}:}~\cdot~e^{:F:}~=~e^{:{{h}}:}
\end{eqnarray}
with
\begin{eqnarray}
f_{2}~=~ -\frac{\mu}{2} \left( \frac{x^{2}}{\beta} + \beta p^{2}_{x}\right),
\end{eqnarray}
where $\mu$ is the overall phase, i.e. the tune $Q$ multiplied by 2$\pi$, and
$\beta$ is the $\beta$-function at the interaction point.
The function $F(x)$ corresponds to the beam--beam potential\footnote{For a discussion of the Lie representation
as a generalized kick map, see \cite{chao}.}
\begin{eqnarray}
F(x) = \displaystyle{\int_{0}^{x}} {\mathrm d}x' f(x').
\end{eqnarray}
For a Gaussian beam we use for $f(x)$ the well-known expression for round beams:
\begin{eqnarray}
f(x) = \frac{2 N r_{0}}{\gamma x} \left( 1 - \mathrm{e}^{\frac{-x^{2}}{2\sigma^{2}}}\right).
\end{eqnarray}
Here $N$ is the number of particles per bunch, $r_{0}$ the classical particle radius,
$\gamma$ the relativistic parameter and $\sigma$ the transverse beam size.

For the analysis we examine the invariant $h$ which determines the one-turn map (OTM) written
as a Lie transformation $e^{:{{h}}:}$.
The invariant $h$ is the effective Hamiltonian for this problem.

As usual we transform to action and angle variables $A$ and $\Phi$, related to the
variables $x$ and $p_{x}$ through the transformations
\begin{eqnarray}
x = \sqrt{2A\beta} \mathrm{sin}(\Phi), ~~~~~p_{x} = \sqrt{\frac{2A}{\beta}}\mathrm{cos}(\Phi).
\end{eqnarray}
With this transformation we get a simple representation
for the linear transfer map $f_{2}$:
\begin{eqnarray}
f_{2} = -\mu A.
\end{eqnarray}
The function $F(x)$ we write as Fourier series:
\begin{eqnarray}
F(x) \Rightarrow \sum_{n=-\infty}^{\infty}  c_{n}(A) \mathrm{e}^{\mathrm{i}n\Phi}
\label{eq:cn}
\end{eqnarray}
with the coefficients $c_{n}(A)$:
\begin{eqnarray}
 c_n(A)~=~\frac{1}{2\pi}\int_{0}^{2\pi} \mathrm{e}^{-\mathrm{i}n\Phi} F(x) {\mathrm d}\Phi.
\label{eq:cn1}
\end{eqnarray}
For the evaluation of (\ref{eq:cn1}), see \cite{chao}.
We take some useful properties of Lie operators (see any textbook, e.g. \cite{ef1, chao}):
\begin{eqnarray}
:f_{2}:g(A) = 0,~~~~~:f_{2}:\mathrm{e}^{\mathrm{i}n\Phi} = \mathrm{i}n \mu \mathrm{e}^{\mathrm{i}n\Phi},~~~~~g(:f_{2}:)\mathrm{e}^{\mathrm{i}n \Phi} = g(\mathrm{i}n \mu) \mathrm{e}^{\mathrm{i}n\Phi}
\end{eqnarray}
and the CBH formula for the concatenation of the maps (see any textbook, e.g. \cite{ef1, chao}):
\begin{eqnarray}
\mathrm{e}^{:f_{2}:}~\mathrm{e}^{:F:}~=~\mathrm{e}^{:h:}~=~\mathrm{exp}\left[:f_{2} + \left( \frac{:f_{2}:}{1 - \mathrm{e}^{-:f_{2}:}}\right) F + {\cal{O}}(F^{2}): \right],
\end{eqnarray}
which gives immediately for $h$
\begin{eqnarray*}
h = -\mu A + \sum_{n}  c_{n}(A) \frac{\mathrm{i} n \mu}{1 - \mathrm{e}^{-\mathrm{i} n \mu}} \mathrm{e}^{\mathrm{i}n\Phi}
\end{eqnarray*}
\begin{eqnarray}
{\mathrm{and}}~~~~~h = -\mu A + \sum_{n}  c_{n}(A) \frac{n \mu}{2 \mathrm{sin}(\frac{n \mu}{2})} \mathrm{e}^{(\mathrm{i}n\Phi + \mathrm{i}\frac{n \mu}{2})}.
\label{eq:h1}
\end{eqnarray}
Away from resonances, a normal-form transformation gives
\begin{eqnarray}
h~=~-\mu A + c_{0}(A)~=\mathrm{const}.
\end{eqnarray}
On resonance, i.e. for the condition
\begin{eqnarray}
Q = \frac{p}{n} = \frac{\mu}{2 \pi}
\end{eqnarray}
and, with $c_{n}~\neq~0$, we have
\begin{eqnarray*}
\mathrm{sin}\left(\frac{n \pi p}{n}\right)  = \mathrm{sin}(p \pi) \equiv 0 ~~~~\mathrm{for~~all~~integers}~~~p
\end{eqnarray*}
and the invariant $h$ diverges.
This is a well-known result and not surprising.

\subsubsection{Non-linear beam--beam tune shift}
Having derived the effective Hamiltonian,
\begin{eqnarray}
h~=~-\mu A + c_{0}(A)~=\mathrm{const.},
\end{eqnarray}
we can now easily write an expression for the non-linear beam--beam tune shift
derived earlier:
\begin{eqnarray}
\Delta Q~=~\frac{\mathrm{d} c_{0}(A)}{\mathrm{d}A}.
\end{eqnarray}
Using the force for a round Gaussian beam and action-angle variables, we write the
beam--beam potential $F(x)$ as
\begin{eqnarray}
F(x) = \frac{N r_{0}}{\gamma} \int_{0}^{A\beta/2\sigma^{2}} (1 - \mathrm{e}^{-2\alpha \sin^{2}(\Phi)})\frac{{\mathrm d} \alpha}{\alpha}.
\end{eqnarray}
The coefficients $c_{n}(A)$ become
\begin{eqnarray}
c_{n}(A) = \frac{N r_{0}}{\gamma} \int_{0}^{A\beta/2\sigma^{2}} \frac{{\mathrm d}\alpha}{\alpha}\frac{1}{2\pi}\int_{0}^{2\pi} {\mathrm d}\Phi \mathrm{e}^{-\mathrm{i}n\Phi}(1 - \mathrm{e}^{-2\alpha \sin^{2}(\Phi)}).
\end{eqnarray}
With the coefficient $c_{0}(A)$ we get for the tune shift as a function of the amplitude:
\begin{eqnarray}
\Delta Q~=~\frac{1}{2\pi}\frac{N r_{0}}{\gamma}\frac{\mathrm{d}}{\mathrm{d}A} \int_{0}^{A\beta/2\sigma^{2}} \frac{{\mathrm d} \alpha}{\alpha} (1 - \mathrm{e}^{-\alpha}I_{0}(\alpha))
\end{eqnarray}
\begin{eqnarray}
=~\frac{1}{2\pi}\frac{N r_{0}}{\gamma A} (1 - I_{0}(A\beta/2\sigma^{2}) \cdot \mathrm{e}^{-A\beta/2\sigma^{2}}),
\end{eqnarray}
which, for $J = A\beta/\sigma^{2}$, is the result we obtained earlier.

\subsubsection{Two interaction points}
\begin{figure}[htb]
\centering{
\includegraphics*[width= 9.2cm, height= 5.2cm]{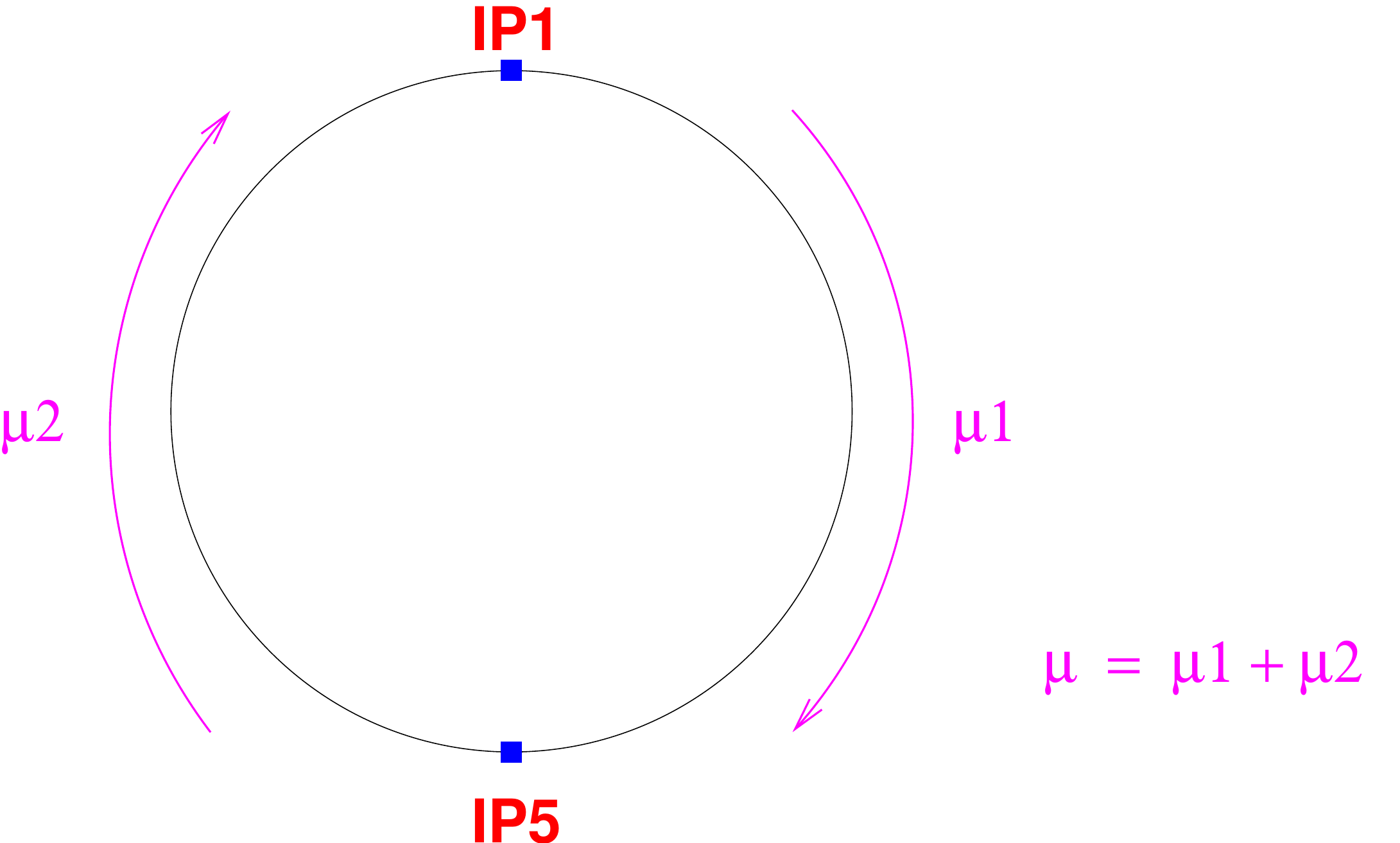}
}
\caption{Two collision points with unequal phase advance}
\label{fig:a02}
\end{figure}

To study two interaction points we use a configuration as shown in Fig.~\ref{fig:a02} and
extend the treatment of a single beam--beam interaction in \cite{chao1} to any number
of beam--beam interactions.
Following the LHC conventions we label the interaction points IP1 and IP5.
The phase advance between IP1 and IP5
is $\mu_{1}$, from IP5 to IP1 it is $\mu_{2}$
and the overall phase advance for one turn is $\mu~=~\mu_{1} + \mu_{2}$.

For the computation we have now two transfers $f_{2}^{1}, f_{2}^{2}$ and two beam--beam
kicks $F^{1}, F^{2}$,
assuming the first interaction point (IP5) at $\mu_{1}$ and the second (IP1) at $\mu$ \cite{dk1}.
The one-turn map with the four transforms is then
\begin{eqnarray}
&= &\mathrm{e}^{:f_{2}^{1}:}~\mathrm{e}^{:F^{1}:}~\mathrm{e}^{:f_{2}^{2}:}~\mathrm{e}^{:F^{2}:}~=~\mathrm{e}^{:h_{2}:}.
\end{eqnarray}

This we can easily re-write using the properties of Lie operators as
\begin{eqnarray*}
&= &\mathrm{e}^{:f_{2}^{1}:}~\mathrm{e}^{:F^{1}:}~\mathrm{e}^{-:f_{2}^{1}:}~\mathrm{e}^{:f_{2}^{1}:}~\mathrm{e}^{:f_{2}^{2}:}~\mathrm{e}^{:F^{2}:}~=~\mathrm{e}^{:h_{2}:}\\
&= &\mathrm{e}^{:f_{2}^{1}:}~\mathrm{e}^{:F^{1}:}~\mathrm{e}^{-:f_{2}^{1}:}~\mathrm{e}^{:f_{2}:}~\mathrm{e}^{:F^{2}:}~\mathrm{e}^{-:f_{2}:}~\mathrm{e}^{:f_{2}:}=~\mathrm{e}^{:h_{2}:}\\
&= &\mathrm{e}^{:\mathrm{e}^{-:f_{2}^{1}:}F^{1}:}~\mathrm{e}^{:\mathrm{e}^{-:f_{2}:}F^{2}:}~\mathrm{e}^{:f_{2}:}=~\mathrm{e}^{:h_{2}:}.\\
\end{eqnarray*}
Assuming now the simplification
\begin{eqnarray}
f_{2}~=~-\mu A,~~~f_{2}^{1}~=~-\mu_{1} A~~~ \mathrm{and} ~~~f_{2}^{2}~=~-\mu_{2} A
\end{eqnarray}
and, remembering that $g(:f_{2}:)\mathrm{e}^{\mathrm{i}n \Phi} = g(\mathrm{i}n \mu) \mathrm{e}^{\mathrm{i}n\Phi}$, we have
\begin{eqnarray}
\mathrm{e}^{:f_{2}^{1}:}\mathrm{e}^{\mathrm{i}n\Phi}~=~\mathrm{e}^{\mathrm{i}n \mu_{1}} \mathrm{e}^{\mathrm{i}n \Phi}~=~\mathrm{e}^{\mathrm{i}n(\mu_{1} + \Phi)}
\end{eqnarray}
and we find that
the Lie transforms of the perturbations are phase shifted (see e.g. \cite{ef1}). Therefore,
\begin{eqnarray}
\mathrm{e}^{:\mathrm{e}^{-:f_{2}^{1}:}F^{1}:}~\mathrm{e}^{:\mathrm{e}^{-:f_{2}:}F^{2}:}~\mathrm{e}^{:f_{2}:}=~\mathrm{e}^{:h_{2}:}
\end{eqnarray}
becomes simpler with substitutions of {{${{\Phi_{1}}}~\rightarrow~\Phi + \mu_{1}$}} and {{${{\Phi}}~\rightarrow~\Phi + \mu$}} in the functions $G^{1}$ and $G$:
\begin{eqnarray}
 &\mathrm{e}^{:G^{1}({{\Phi_{1}}}):} \mathrm{e}^{:G({{\Phi}}):} \mathrm{e}^{:f_{2}:}~~~{{\Rightarrow}}~~~
 &\mathrm{e}^{:G^{1}({{\Phi_{1}}}) + G({{\Phi}}):} \mathrm{e}^{:f_{2}:}
\end{eqnarray}
This reflects the phase-shifted distortions and we get for $h_{2}$
\begin{eqnarray}
h_{2} = -\mu A + \sum_{n=-\infty}^{\infty}  \frac{n \mu c_{n}(A)}{2\mathrm{sin}(n\frac{\mu}{2})} \left[{\mathrm{e}^{-\mathrm{i}n(\Phi+\mu/2 + \mu_{1})}} + {\mathrm{e}^{-\mathrm{i} n(\Phi+\mu/2)}}\right]
\end{eqnarray}
or, re-written,\footnote{For head-on collisions only $c_{n}(A)$ for even orders in $n$ are non-zero and the sum needs to be done only for even terms}:
\begin{eqnarray}
h_{2} = -\mu A + 2c_{0}(A) + {\underbrace{\sum_{n=2}^{\infty}  \frac{2 n \mu c_{n}(A)}{\mathrm{sin}(n\frac{\mu}{2})} \mathrm{cos}\left[n(\Phi + \frac{\mu}{2} + \frac{\mu_{1}}{2})\right] \mathrm{cos}(n \frac{\mu_{1}}{2})}_{\mathrm{interesting~part}}}.
\label{eq:h2}
\end{eqnarray}
Note well, because
\begin{eqnarray}
 \mathrm{e}^{:F(\Phi):} \mathrm{e}^{:f_{2}:}~~\rightarrow~~\mathrm{e}^{:G^{1}({{\Phi_{1}}}) + G({{\Phi}}):} \mathrm{e}^{:f_{2}:},
\end{eqnarray}
that the above treatment can be generalized to more interaction points, in particular including long-range
interactions.

In practice, Eq. (\ref{eq:h2}) is evaluated to a maximum order $N$, in our case up to order 40.
We get
\begin{eqnarray}
h_{2} = -\mu A + 2c_{0}(A) + {\underbrace{\sum_{n=2}^{N}  \frac{2 n \mu c_{n}(A)}{\mathrm{sin}(n\frac{\mu}{2})} \mathrm{cos}\left[n(\Phi + \frac{\mu}{2} + \frac{\mu_{1}}{2})\right] \mathrm{cos}(n \frac{\mu_{1}}{2})}_{\mathrm{interesting~part}}}
\label{eq:h2a}
\end{eqnarray}
with $N$ = 40.

\newpage
\subsubsection{Comparison with numerical model}
To test our result, we compare the invariant $h$ to the
results of a particle-tracking program \cite{dk1}.

The model we use in the program is rather simple:
\begin{itemize}
\item[$\bullet$] linear transfer between interactions;
\item[$\bullet$] beam--beam kick for round beams;
\item[$\bullet$] compute action $I~=~\frac{\beta^{*}}{2 \sigma^{2}}(\frac{x^{2}}{\beta^{*}} + p_{x}^{2}\beta^{*})$;
\item[$\bullet$] compute phase $\Phi~=~{\mathrm{arctan}}(\frac{p_{x}}{x})$;
\item[$\bullet$] compare $I$ with $h$ as a function of the phase $\Phi$.
\end{itemize}
The evaluation of the invariant (\ref{eq:h1}) is done numerically with Mathematica.

\begin{figure}[htb]
\centering{
\includegraphics*[width= 5.2cm, height= 3.5cm]{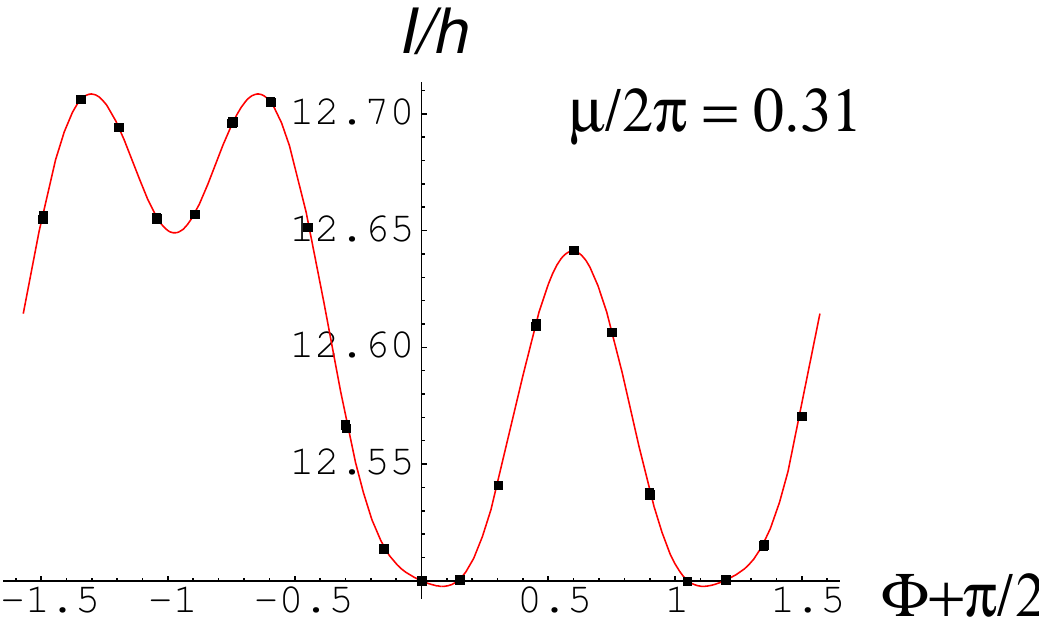}
\includegraphics*[width= 5.2cm, height= 3.5cm]{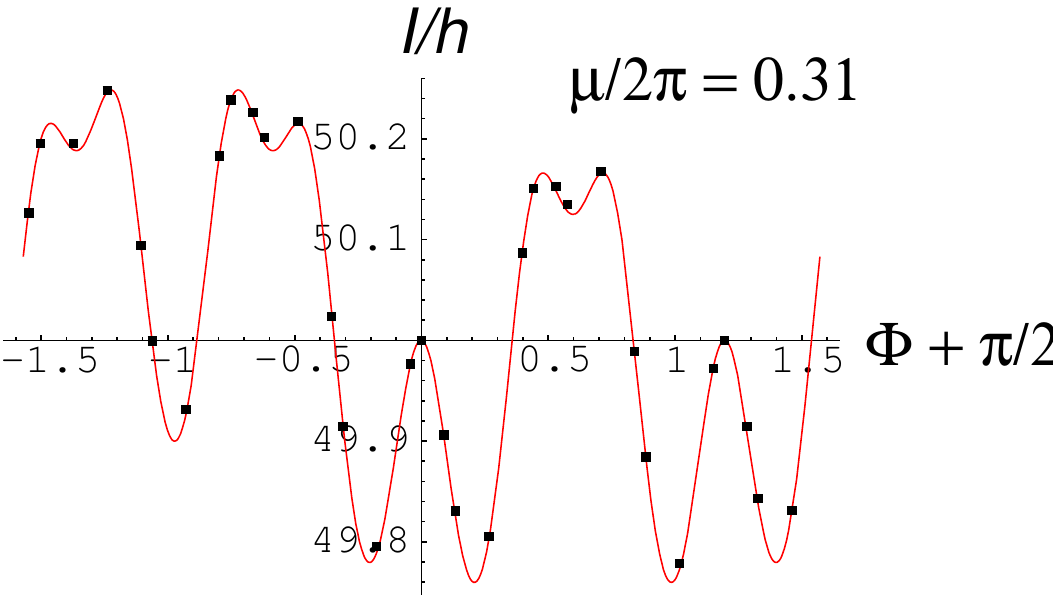}}
\caption{Comparison: numerical and analytical models for one interaction point. Shown for 5$\sigma_{x}$ and 10$\sigma_{x}$. Full symbols from numerical model and solid lines from invariant (\ref{eq:h1}).}
\label{fig:a04}
\end{figure}

The comparison between the tracking results and the invariant $h$ from the analytical calculation
is shown in Fig.~\ref{fig:a04} in the ($I$, $\Phi$) space.
Only one interaction point is used in this comparison and the particles are tracked for 1024 turns.
The symbols are the results from the tracking and the solid lines are the invariants computed as above.
The two figures are computed for amplitudes of 5$\sigma$ and 10$\sigma$.
The agreement between the models is excellent.
The analytical calculation was again done up to the order $N = 40$.
Using a lower number, the analytical model can reproduce the envelope of the tracking
results, but not the details.

\begin{figure}[htb]
\centering{
\includegraphics*[width= 5.2cm, height= 3.5cm]{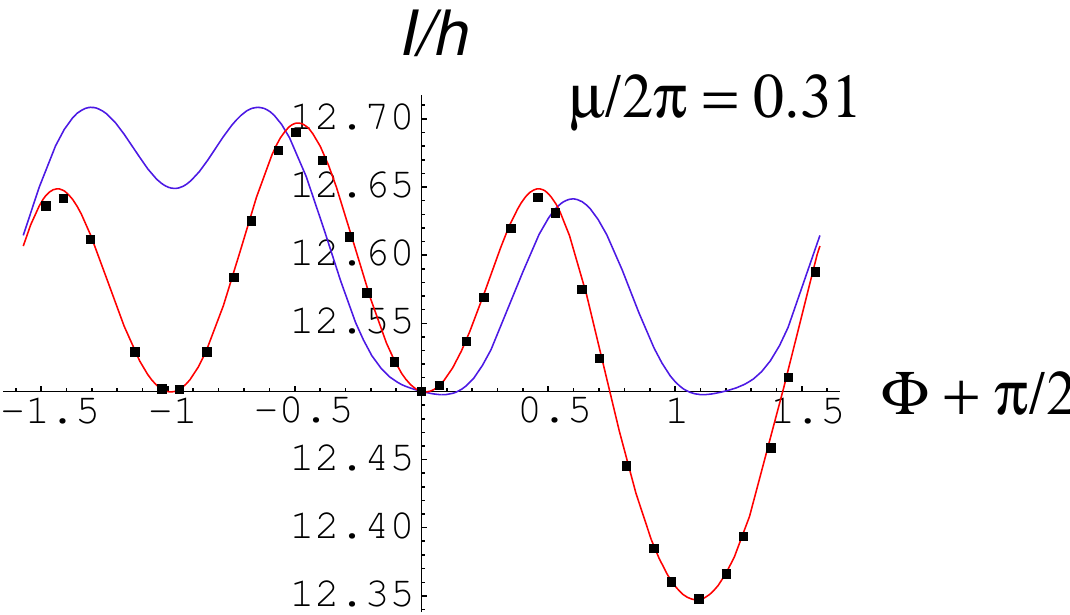}
\includegraphics*[width= 5.2cm, height= 3.5cm]{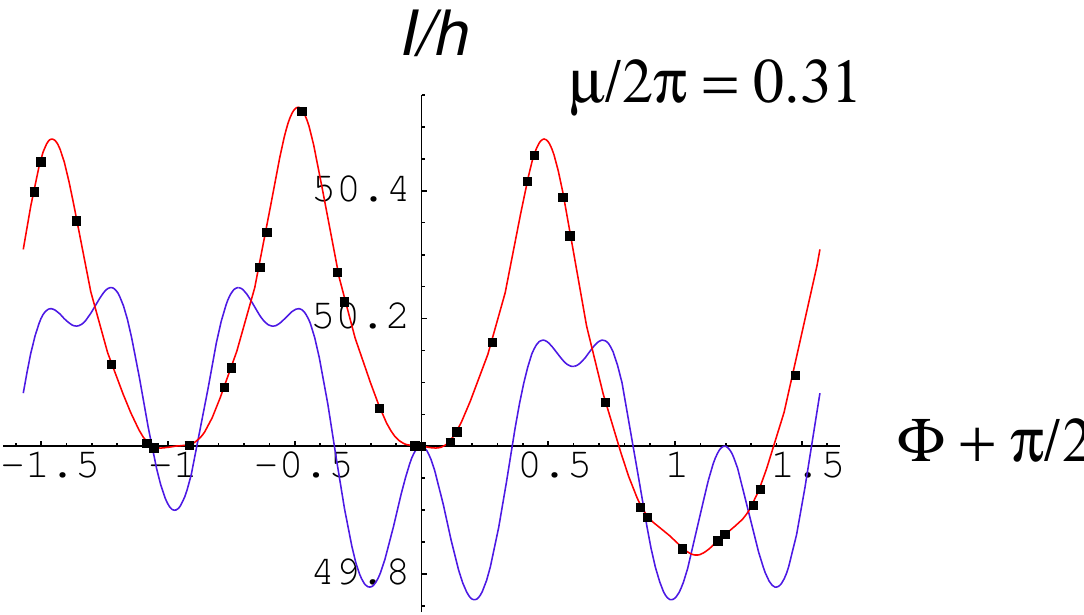}
}
\caption{Comparison: numerical and analytical models for two interaction points.
Shown for 5$\sigma_{x}$ and 10$\sigma_{x}$.
Full symbols from numerical model and solid lines from invariant.
Shown here are the invariant for one (Eq. (\ref{eq:h1}), blue line, not passing through the full symbols) and two (Eq. (\ref{eq:h2}), red line, passing through the full symbols) interactions to demonstrate the
difference and the agreement with the tracking program.}
\label{fig:a05}
\end{figure}

Another comparison is done in Fig.~\ref{fig:a05} for the case of two interaction points.
The symbols are again from the simulation and the solid lines from the computation.
For comparison, the invariant for a single interaction point is included to demonstrate
the difference.
Again the agreement is excellent and shows the validity of the results.

\subsection{Behaviour near a resonance}
\begin{figure}[htb]
\centering{
\includegraphics*[width= 5.2cm, height= 3.5cm]{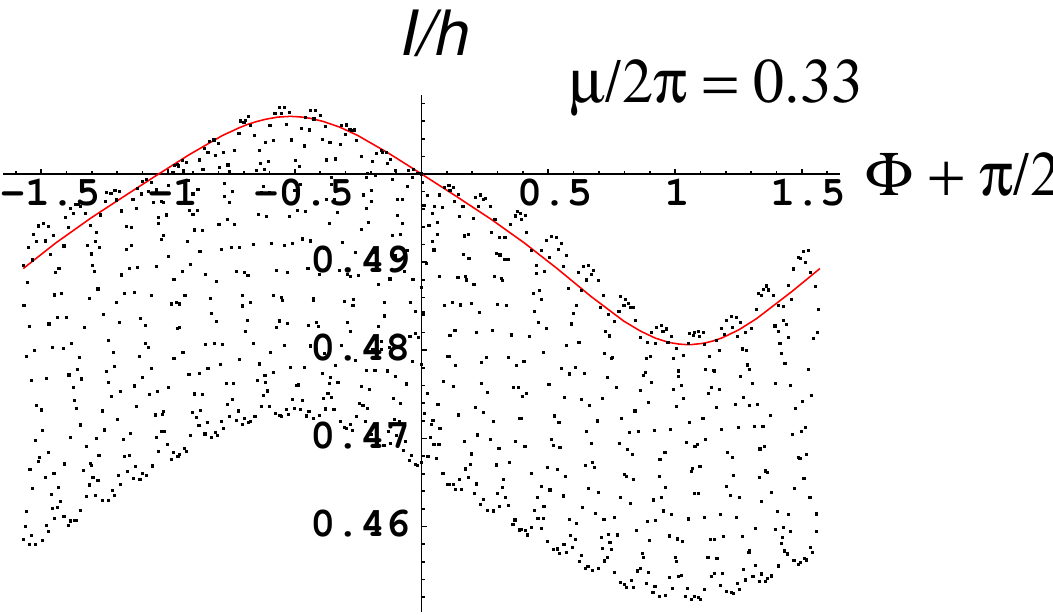}
\includegraphics*[width= 5.2cm, height= 3.5cm]{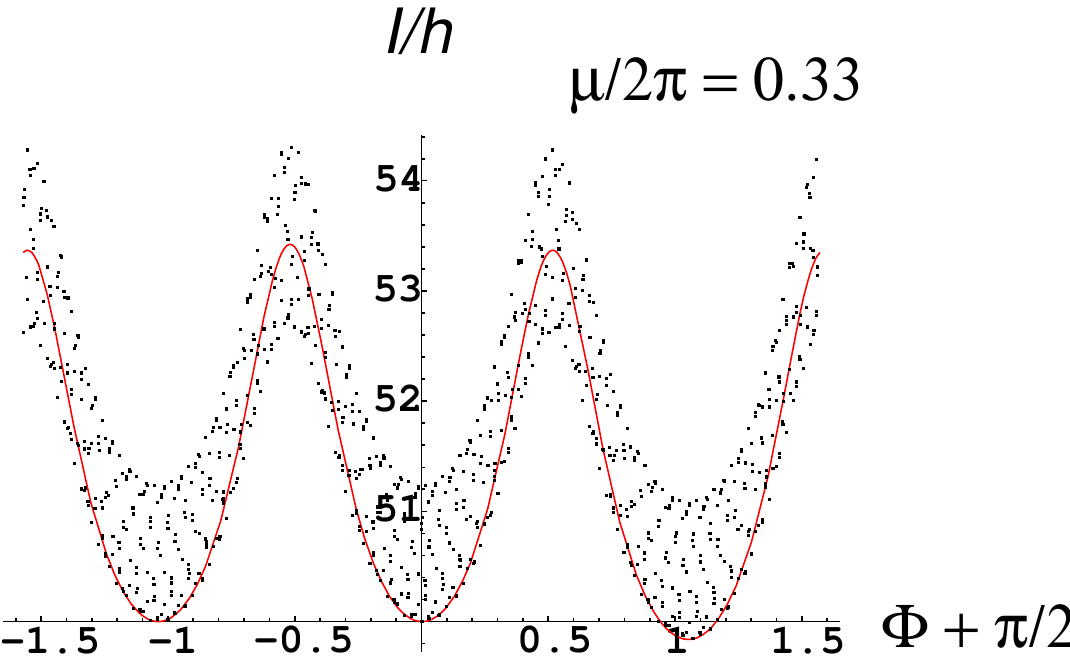}
}
\caption{Comparison: numerical and analytical models for two interaction points on a resonance (third order)}
\label{fig:06}
\end{figure}

To show the behaviour of the system near a resonance,
we show the invariant together with the tracking results near the third-order
resonance in Fig.~\ref{fig:06}.

It is clearly demonstrated that the simulation differs quantitatively from the computed
invariant at resonant tunes.

\end{document}